\documentclass[
 reprint,
 nofootinbib,
 amsmath,amssymb,
 aps,
 superscriptaddress
]{revtex4-2}

\usepackage{comment}
\usepackage{graphicx}
\usepackage{dcolumn}
\usepackage{bm}
\usepackage{tabularray}

\usepackage{hyperref}
\hypersetup{
    colorlinks=true,
    linkcolor=blue,
    filecolor=magenta,      
    urlcolor=blue,
    citecolor=blue
    }

\begin{document}

\preprint{APS/123-QED}

\title{Reaction plane correlated triangular flow in Au+Au collisions at $\sqrt{s_{NN}}=3$ GeV}

\affiliation{Abilene Christian University, Abilene, Texas   79699}
\affiliation{AGH University of Krakow, FPACS, Cracow 30-059, Poland}
\affiliation{Argonne National Laboratory, Argonne, Illinois 60439}
\affiliation{American University in Cairo, New Cairo 11835, Egypt}
\affiliation{Ball State University, Muncie, Indiana, 47306}
\affiliation{Brookhaven National Laboratory, Upton, New York 11973}
\affiliation{University of Calabria \& INFN-Cosenza, Rende 87036, Italy}
\affiliation{University of California, Berkeley, California 94720}
\affiliation{University of California, Davis, California 95616}
\affiliation{University of California, Los Angeles, California 90095}
\affiliation{University of California, Riverside, California 92521}
\affiliation{Central China Normal University, Wuhan, Hubei 430079 }
\affiliation{University of Illinois at Chicago, Chicago, Illinois 60607}
\affiliation{Creighton University, Omaha, Nebraska 68178}
\affiliation{Czech Technical University in Prague, FNSPE, Prague 115 19, Czech Republic}
\affiliation{Technische Universit\"at Darmstadt, Darmstadt 64289, Germany}
\affiliation{National Institute of Technology Durgapur, Durgapur - 713209, India}
\affiliation{ELTE E\"otv\"os Lor\'and University, Budapest, Hungary H-1117}
\affiliation{Frankfurt Institute for Advanced Studies FIAS, Frankfurt 60438, Germany}
\affiliation{Fudan University, Shanghai, 200433 }
\affiliation{Guangxi Normal University, Guilin, 541004 }
\affiliation{University of Heidelberg, Heidelberg 69120, Germany }
\affiliation{University of Houston, Houston, Texas 77204}
\affiliation{Huzhou University, Huzhou, Zhejiang  313000}
\affiliation{Indian Institute of Science Education and Research (IISER), Berhampur 760010 , India}
\affiliation{Indian Institute of Science Education and Research (IISER) Tirupati, Tirupati 517507, India}
\affiliation{Indian Institute Technology, Patna, Bihar 801106, India}
\affiliation{Indiana University, Bloomington, Indiana 47408}
\affiliation{Institute of Modern Physics, Chinese Academy of Sciences, Lanzhou, Gansu 730000 }
\affiliation{University of Jammu, Jammu 180001, India}
\affiliation{Kent State University, Kent, Ohio 44242}
\affiliation{University of Kentucky, Lexington, Kentucky 40506-0055}
\affiliation{Lawrence Berkeley National Laboratory, Berkeley, California 94720}
\affiliation{Lehigh University, Bethlehem, Pennsylvania 18015}
\affiliation{Max-Planck-Institut f\"ur Physik, Munich 80805, Germany}
\affiliation{Michigan State University, East Lansing, Michigan 48824}
\affiliation{National Institute of Science Education and Research, HBNI, Jatni 752050, India}
\affiliation{National Cheng Kung University, Tainan 70101 }
\affiliation{Nuclear Physics Institute of the CAS, Rez 250 68, Czech Republic}
\affiliation{The Ohio State University, Columbus, Ohio 43210}
\affiliation{Institute of Nuclear Physics PAN, Cracow 31-342, Poland}
\affiliation{Panjab University, Chandigarh 160014, India}
\affiliation{Purdue University, West Lafayette, Indiana 47907}
\affiliation{Rice University, Houston, Texas 77251}
\affiliation{Rutgers University, Piscataway, New Jersey 08854}
\affiliation{Universidade de S\~ao Paulo, S\~ao Paulo, Brazil 05314-970}
\affiliation{University of Science and Technology of China, Hefei, Anhui 230026}
\affiliation{South China Normal University, Guangzhou, Guangdong 510631}
\affiliation{Sejong University, Seoul, 05006, South Korea}
\affiliation{Shandong University, Qingdao, Shandong 266237}
\affiliation{Shanghai Institute of Applied Physics, Chinese Academy of Sciences, Shanghai 201800}
\affiliation{Southern Connecticut State University, New Haven, Connecticut 06515}
\affiliation{State University of New York, Stony Brook, New York 11794}
\affiliation{Instituto de Alta Investigaci\'on, Universidad de Tarapac\'a, Arica 1000000, Chile}
\affiliation{Temple University, Philadelphia, Pennsylvania 19122}
\affiliation{Texas A\&M University, College Station, Texas 77843}
\affiliation{University of Texas, Austin, Texas 78712}
\affiliation{Tsinghua University, Beijing 100084}
\affiliation{University of Tsukuba, Tsukuba, Ibaraki 305-8571, Japan}
\affiliation{University of Chinese Academy of Sciences, Beijing, 101408}
\affiliation{United States Naval Academy, Annapolis, Maryland 21402}
\affiliation{Valparaiso University, Valparaiso, Indiana 46383}
\affiliation{Variable Energy Cyclotron Centre, Kolkata 700064, India}
\affiliation{Warsaw University of Technology, Warsaw 00-661, Poland}
\affiliation{Wayne State University, Detroit, Michigan 48201}
\affiliation{Wuhan University of Science and Technology, Wuhan, Hubei 430065}
\affiliation{Yale University, New Haven, Connecticut 06520}

\author{M.~I.~Abdulhamid}\affiliation{American University in Cairo, New Cairo 11835, Egypt}
\author{B.~E.~Aboona}\affiliation{Texas A\&M University, College Station, Texas 77843}
\author{J.~Adam}\affiliation{Czech Technical University in Prague, FNSPE, Prague 115 19, Czech Republic}
\author{L.~Adamczyk}\affiliation{AGH University of Krakow, FPACS, Cracow 30-059, Poland}
\author{J.~R.~Adams}\affiliation{The Ohio State University, Columbus, Ohio 43210}
\author{I.~Aggarwal}\affiliation{Panjab University, Chandigarh 160014, India}
\author{M.~M.~Aggarwal}\affiliation{Panjab University, Chandigarh 160014, India}
\author{Z.~Ahammed}\affiliation{Variable Energy Cyclotron Centre, Kolkata 700064, India}
\author{E.~C.~Aschenauer}\affiliation{Brookhaven National Laboratory, Upton, New York 11973}
\author{S.~Aslam}\affiliation{Indian Institute Technology, Patna, Bihar 801106, India}
% ...

\collaboration{STAR Collaboration}\noaffiliation

\date{April 19, 2024}

\begin{abstract}
We measure triangular flow relative to the reaction plane at 3 GeV center-of-mass energy in Au+Au collisions at the BNL Relativistic Heavy Ion Collider. A significant $v_3$ signal for protons is observed, which increases for higher rapidity, higher transverse momentum, and more peripheral collisions.  The triangular flow is essentially rapidity-odd with a slope at mid-rapidity, $dv_3/dy|_{(y=0)}$, opposite in sign compared to the slope for directed flow. No significant $v_3$ signal is observed for charged pions and kaons.
Comparisons with models suggest that a mean field potential is required to describe these results, and that the triangular shape of the participant nucleons is the result of stopping and nuclear geometry. 
\end{abstract}

\maketitle

\section{\label{sec:level1}Introduction}
One of the primary objectives of the Beam Energy Scan II (BES II), undertaken by the STAR collaboration \cite{Odyniec:2019kfh}, is to identify and study the transition from hadronic matter to the Quark Gluon Plasma (QGP). This phase transition, believed to be first-order at high baryon density \cite{Muller:2022qsm}, requires a comprehensive understanding of nuclear matter at extreme conditions. A critical aspect of this research is to discern how nuclear matter transforms from a state of high baryon density nucleons in low-energy heavy ion collisions to a QGP in higher energy collisions. BES II aims to advance our knowledge by investigating anisotropic flow in Au+Au collisions at low collision energies and progressively examining higher energies to observe the evolution of flow as the phase transition is approached. This paper presents the results of anisotropic flow measurements, specifically triangular flow as described below, at $\sqrt{s_{NN}}=3.0$ GeV, the lowest energy in the BES II program at STAR.

Anisotropic flow describes the shape and direction of expansion of the medium produced in heavy-ion collisions. As an observable, flow manifests itself in the azimuthal particle distribution relative to the true reaction plane $\Psi_r$. This is expressed mathematically in the following triple differential distribution expanded as a Fourier series that describes the distribution of final state particles \cite{Poskanzer}:
\begin{equation}
    E\frac{d^3N}{d^3p}=\frac{1}{2\pi}\frac{d^2N}{p_{\mathrm{T}}dp_{\mathrm{T}}dy}(1+\sum_{n=1}^\infty 2v_n \cos(n(\phi-\Psi_r))),
\end{equation}
where $p_{\mathrm{T}}, y, \phi,$ and $\Psi_r$ are the the particle transverse momentum, rapidity, azimuthal angle, and the true reaction plane angle, respectively. The coefficients in the expansion, $v_1$ (directed flow), $v_2$ (elliptic flow), $v_3$ (triangular flow), etc., describe the collective response of the medium to the shape of the initial collision geometry. They are sensitive to medium properties such as the viscosity and mean-fields that determine the equation of state (EOS).

Anisotropic collective flow has been extensively studied in heavy-ion collisions, where early studies were carried out at the Bevelac and at GSI (see \cite{Reisdorf:1997fx} for a review) which studied primarily $v_1$ and $v_2$ at $\sqrt{s_{NN}} \approx$ a few hundred MeV. Later experiments followed at the AGS ($\sqrt{s_{NN}} \approx 5$ GeV), and the SPS ($\sqrt{s_{NN}} \approx 20$ GeV). A notable observation of collisions at energies $\sqrt{s_{NN}} \lesssim 4$ GeV was that the overlapping nucleons (participants) could begin flowing outward before the outside, non-overlapping nucleons (spectators), could fully separate from the participants. This blocking of flow by spectators is known as shadowing which leads to phenomena such as in-plane and out-of-plane flow \cite{E895:1999ldn}. 

The timescale of the collision between the nuclei is a critical element, both in the creation of the medium and in the formation of its shape. At high energies, above $\sqrt{s_{NN}}$ = 27 GeV, the transit time, $\tau \approx 2R/\gamma\beta$ (where $R$ is the radius of the nucleus, $\gamma$ is the Lorentz factor, and $\beta$ is the velocity of the nuclei) is much shorter than the formation time of particles. In these cases, the spectators are well away from the collision volume and the medium is free to expand \cite{Lin:2017lcj}.
Another observation at these energies is that a non-zero $v_3$ develops that is caused by the participants randomly arranging into a triangle shape in some events. These triangular shapes can occur at any angle with respect to $\Psi_{r}$, so there is no correlation between $v_3$ and $\Psi_{r}$ (the correlation is only with its associated third-order event plane angle $\Psi_3$). This $v_3$ is rapidity even, with a magnitude that decreases towards lower $\sqrt{s_{NN}}$. It has been studied by various experiments, in particular by the STAR collaboration which has reported values of $p_{\textrm{T}}$ integrated $v_3$ of $\approx$ 2\% at $\sqrt{s_{NN}}$ = 200 GeV down to $\approx$ 1\% at 7.7 GeV \cite{STAR:2016vqt}.

This paper reports the observation of $v_3$ that is \it{correlated} with the reaction plane.\rm \footnote[1]{The term ``reaction plane'' (as opposed to ``event plane'') is deliberately used here to describe the first-order event plane since it should be an approximation of $\Psi_{r}$ and also to avoid confusing it with the third-order event plane that is usually associated with triangular flow.} We denote this observable as $v_3\{\Psi_1\}$ ($v_3$ calculated with the first-order event plane) to distinguish it from the fluctuation-driven $v_3$ discussed above at higher energies. The measurement of $v_3\{\Psi_1\}$ has also been made by the HADES collaboration in Au+Au collisions at $\sqrt{s_{NN}}=2.4$ GeV \cite{HADES:2020lob, Kardan:2017qau}. Our study measures $v_3\{\Psi_1\}$ at comparatively higher energy and introduces comparisons to theoretical models, to gain insight into the origin of this phenomenon. 

\begin{figure}
    \centering
    \includegraphics[scale=0.21]{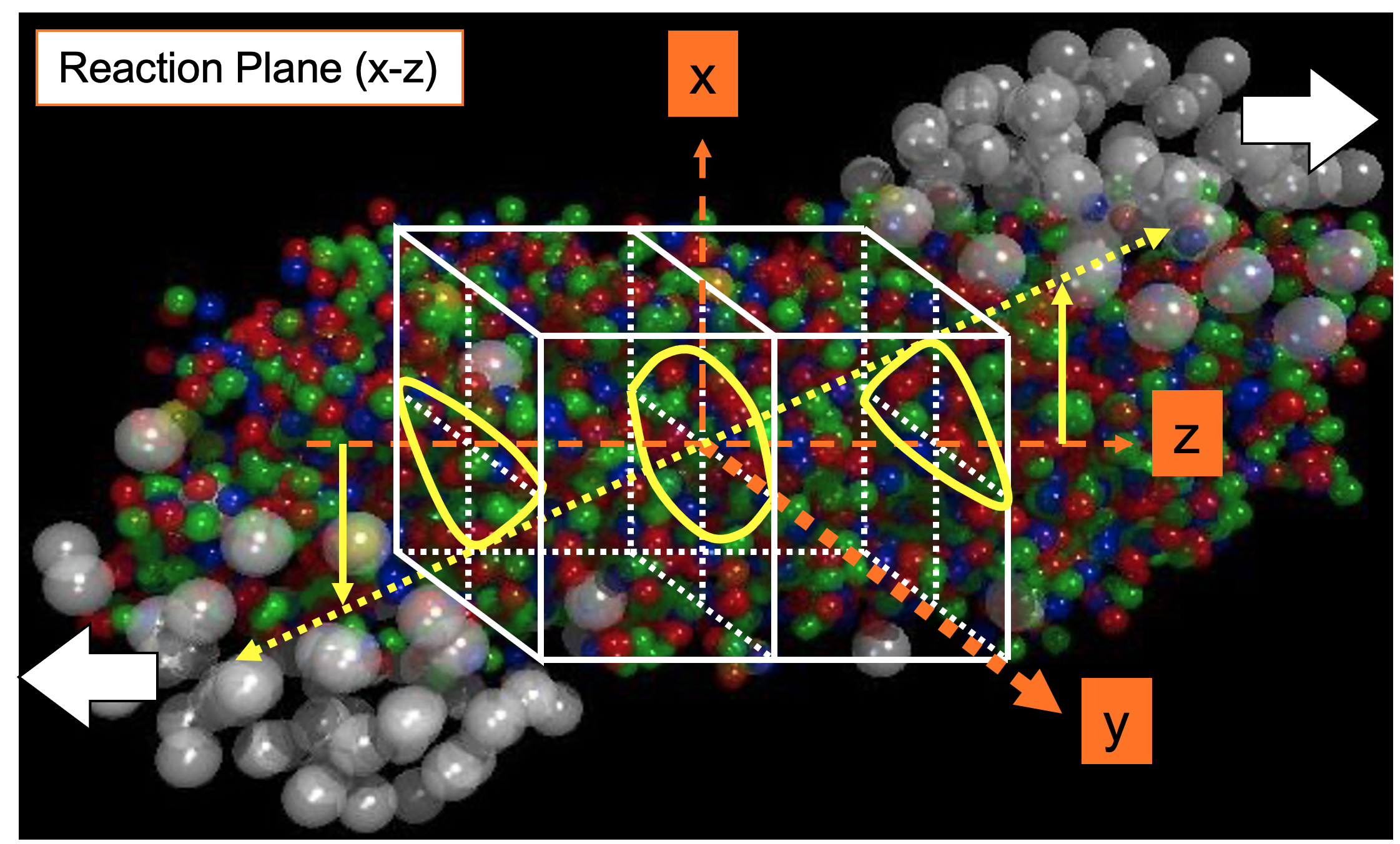}
    \caption{Cartoon of a noncentral heavy-ion collision illustrating the triangular shape correlated with the event plane (the $x$-$z$ plane). The forces on the particles resulting from the triangular geometry are perpendicular to the sides of the yellow triangles, opposite to the direction of $v_1$. Note that the triangular regions are composed primarily of participants. }
    \label{fig:true_v3}
\end{figure}

The development of flow typically hinges on two crucial elements. The first pertains to a geometric shape in spatial configuration space that corresponds to a specific flow coefficient, for instance, an almond shape corresponds to $v_2$ and a triangular shape corresponds to $v_3$. 
This paper discusses a mechanism that produces an initial triangular shape which is correlated to the reaction plane and is the result of shadowing and baryon stopping. The nonrandom orientation of this initial triangular shape relative to the reaction plane provides the first necessary element for generating a reaction plane-correlated $v_3$. Figure \ref{fig:true_v3} presents a rudimentary geometric illustration\footnote[2]{While this illustrative model originates from the simulation of higher-energy Au+Au collisions, it is beneficial as it delineates the regions of spectators, participants, and the triangular shape formed by the mechanism outlined in this article.} of the triangular shape situated in both positive and negative rapidity regions, thereby leading to a flow which is odd in rapidity. 
We assume positive rapidity is to the right on the plot, hence $v_1$ is positive \cite{STAR:2021yiu}. The triangle, for example, shown on the right side of the $x$-$y$ plane, is oriented with one edge facing in the negative-$x$ direction, aligning in the direction of the pressure gradient and therefore one direction of flow. The other two major directions would be at 60 degrees to the negative-$x$ direction, hence $v_3\{\Psi_1\}$ will be negative at positive rapidity, opposite to that of $v_1$.

The second element required for the development of the flow described in this paper, are potentials of strong nuclear force associated with the EOS of the produced medium resulting in forces between participants that transform the spacial configuration of particles into a momentum distribution.
At low center-of-mass energies, mean fields provide this force. Flow appears to be sensitive to a variety of parameters that describe these mean fields. Specifically, the incompressibility ($K$) of matter at high baryon density proves to be of particular importance \cite{Nara:2019qfd,Bass:1998ca,Weil:2016zrk}.
In the majority of models, mean fields primarily impact baryons, whereas their effects on mesons originate from interactions or decays that involve baryons \cite{Agnieszka}.
Models that integrate a realistic EOS, which includes a transition from a hadron gas to a QGP, will be indispensable for understanding observations in high energy heavy-ion collisions at high baryon density.

\section{\label{sec:level2}Experiment}

The Solenoidal Tracker at the Relativistic Heavy Ion Collider (RHIC) (STAR) is a multipurpose detector designed to measure hadronic and electromagnetic particles produced in heavy-ion and polarized proton-proton collisions. STAR comprises several subsystems that provide charged particle tracking and identification over a wide range of pseudorapidity ($\eta$) and full azimuth ($\phi$) \cite{Ackermann}. The primary subsystems used for the present analysis are the time projection chamber (TPC) \cite{Anderson}, the barrel time-of-flight detector (TOF) \cite{LLOPE}, and event plane detector (EPD) \cite{Adams}.
The EPD consists of two highly segmented circular detectors positioned at both ends of the TPC and centered around the beam pipe. Each wheel consists of 372 scintillating tiles, enabling a probabilistic determination of the number of minimum ionizing particles (nMIP) passing through each tile per event. The nMIP values are defined within a certainty range of the signal in the scintillator tiles to suppress detector noise and large Landau fluctuations.

In the fixed target (FXT) mode of operation, a single beam strikes a gold foil placed at $z = 200.7$ cm inside the beam pipe on the west side of STAR, near the edge of the TPC. The target has a thickness of 0.25 mm corresponding to a 1$\%$ interaction probability. In this mode, the EPD covers $-5.8 \leq \eta \leq -2.4$ (only the east side can be used), the TPC covers the region $-2 \leq \eta \leq -0.1$, and the TOF covers the range $-1.5 \leq \eta \leq -0.1$ in the laboratory frame; all three subsystems cover the full azimuthal angle.

\section{\label{sec:level3}Analysis}
\subsection{\label{sec:selection}Event and track selection}

The Au+Au data set at $\sqrt{s_{NN}}=3$ GeV was obtained in 2018, with a beam energy of 3.85 GeV per nucleon, in FXT mode. A total of $3.05\times10^8$ events were available for analysis. Midrapidity ($y_{\mathrm{mid}}$) in the laboratory rest frame at this energy is at $y_{\mathrm{mid}}=-1.05$, and center-of-mass rapidity is defined as $y_{\mathrm{c.m.}}=y-y_{\mathrm{mid}}$. This analysis uses the same sign convention of rapidity as in Ref. \cite{STAR:2021beb}, where $y_{\mathrm{c.m.}}<0$ is the forward region and $y_{\mathrm{c.m.}}>0$ is the backward region; hence the beam direction is towards negative rapidity. 

For each event, the reconstructed primary vertex was required to be within 2 cm of the target position along the beam axis. The transverse $x$, $y$ position of the vertex was required to be within a radius of 1.5 cm from the center of the target. These requirements ensured that the event originated from the gold foil and eliminated beam interactions with the vacuum pipe. The event centrality was estimated from the charged particle multiplicity measured in the TPC and categorized into bins of 5\% up to a maximum of 60\%. 
Events with high multiplicity ($>195$ primary tracks) were rejected to avoid pile-up, and events with multiplicity too low ($< 16$ primary tracks) were also rejected.
The remaining pile-up contamination was previously estimated by another STAR analysis at this same energy that studied cumulants of the proton multiplicity distribution. In that analysis, the cumulants were corrected by an unfolding method that statistically separated the single and double collisions in the reconstructed particle multiplicities. This process determined the pile-up fraction to be ($0.46 \pm 0.09$)\% of all events and ($2.10 \pm 0.40$)\% in the 0\%–5\% centrality class \cite{STAR:128.202303}. 

Tracks were required to be reconstructed with at least 15 hit points and greater than 52\% of the total possible points to ensure good track fitting quality. Additionally, the distance of closest approach (DCA) to the primary vertex was required to be less than 3 cm to ensure that the tracks originated from the event vertex. Lastly, we required that at least five measurements for the average energy loss per unit length ($\langle dE/dx \rangle$) were made for each track.

\subsection{\label{sec:particleID}Particle identification}

\begin{figure}
    \centering
    \includegraphics[scale=0.42]{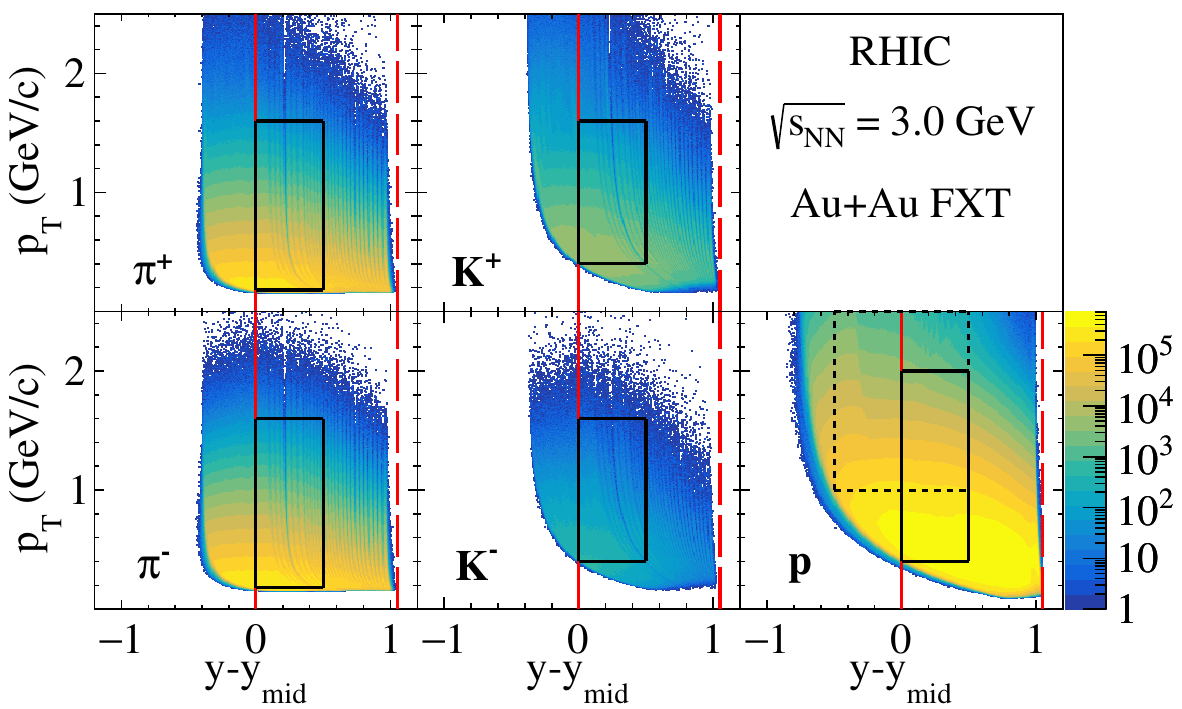}
    \caption{$p_{T}$ vs. $y_{\mathrm{c.m.}}$ density plots for $\pi^{\pm}$, $K^{\pm}$, and $p$ measured by the STAR in Au+Au collisions at $\sqrt{s_{NN}} = 3.0$ GeV. The red dashed line represents the target rapidity and the solid red line represents midrapidity. The solid and dashed black boxes mark acceptance regions used for flow calculations in various cases explained in the text.}
    \label{fig:normalAcceptance}
\end{figure}

$\pi^{\pm}$, $K^{\pm}$, and $p$ in this analysis were identified using a combination of $\langle dE/dx \rangle$ measurements from the TPC and mass information calculated from the time-of-flight provided by the TOF.
Figure \ref{fig:normalAcceptance} shows the $p_{\mathrm{T}}$ and $y_{\mathrm{c.m.}}$ acceptance of each particle type. 
Charged pions and kaons are required to have a $\langle dE/dx \rangle$ measurement within $3\sigma$ of their expected value, and $m^2$ measurements within $-0.1<m^2<0.1$ GeV and $0.15<m^2<0.34$ GeV, respectively. Protons do not require a $m^2$ measurement from the TOF due to their abundance at this collision energy and the fact that, over the relevant momentum range, their $\langle dE/dx \rangle$ curves are well separated from other particles. They are required to have a $\langle dE/dx \rangle$ measurement from the TPC, but we only accept tracks within $2\sigma$ of their expected value to reduce the small contamination that may remain at higher momenta. $\pi^{+}$ and $\pi^{-}$ used for flow measurements were selected within a range of transverse momentum of $0.18 < p_{T} < 1.6$ GeV/$c$, $K^{+}$ and $K^{-}$ within $0.4 < p_{T} < 1.6$ GeV/$c$, and $p$ within $0.4 < p_{T} < 2.0$ GeV/$c$. For flow vs. centrality (and $p_{\mathrm{T}}$), each particle type was selected with a center-of-mass rapidity range of $0 < y_{\mathrm{c.m.}} < 0.5$. These acceptance regions are marked by the solid black boxes in Fig. \ref{fig:normalAcceptance}. For flow measurements vs. $y_{\mathrm{c.m.}}$, the acceptance in $y_{\mathrm{c.m.}}$ is extended for protons to $0 < y_{\mathrm{c.m.}} < 1.0$. Due to the wider range of $y_{\mathrm{c.m.}}$ for protons, we also made a rapidity symmetric acceptance with $-0.5 < y_{\mathrm{c.m.}} < 0.5$ and $1.0 < p_{\mathrm{T}} < 2.5$ GeV/$c$ as shown by the dashed black box in Fig. \ref{fig:normalAcceptance}.

\subsection{\label{sec:EPA}Event plane analysis}
The event plane method for calculating anisotropic flow was employed in this study. This approach utilizes event plane angles ($\Psi_{1}$) reconstructed by a region of the EPD to determine flow coefficients of identified particles in the TPC using
\begin{equation}
    v_3 = \left\langle \frac{\cos(3(\phi-\Psi_1))}{R_{31}} \right\rangle,
\end{equation}
where $R_{31}$ is the resolution correction for the observed event plane angle $\Psi_1$. Here $\phi$ is the azimuthal angle of an identified particle species, and the averages are taken over all tracks of that particle species and all events of a specific centrality class \cite{Poskanzer}. For $\pi^{\pm}$ and $K^{\pm}$, the averages are weighted by the inverse of the product of TPC and TOF matching efficiencies. The averages for protons were weighted by only the inverse of the TPC tracking efficiency. To prevent systematic overestimation of flow we applied the resolution correction track-by-track within the average as suggested in Ref. \cite{MASUI2016181}.

The event plane resolution $R_{31}$ is calculated for each centrality interval using the standard three-subevent method \cite{Poskanzer}:
\begin{equation}\label{eqn:resolution}
     R_{31} = \sqrt{\frac{\langle \cos(3(\Psi^A_1 - \Psi^B_1)) \rangle  \langle \cos(3(\Psi^A_1 - \Psi^C_1)) \rangle}{\langle \cos(3(\Psi^B_1 - \Psi^C_1)) \rangle}},
\end{equation}
where the reaction plane angles $\Psi^A_1$, $\Psi^B_1$, and $\Psi^C_1$ are obtained in three $\eta$ ranges: $-5.8<\eta_A<-3.2$,
$-3.2<\eta_B<-2.5$, and $-1<\eta_C<0$. Equation (\ref{eqn:resolution}) provides the resolution for subevent $A$ which was used for reconstructing $\Psi_1$. Regions $B$ and $C$ were only employed for calculating $R_{31}$, so while the identified particles overlap with region $C$, there is no auto-correlation effect in the flow measurements.

The $\vec{Q}$ vectors used for reconstructing event plane angles are defined as $\vec{Q} = \left(\sum_i w_i cos(n\phi_i), \sum_i w_i sin(n\phi_i)\right)$, where the sums are over all tracks in a particular subevent \cite{Poskanzer}. The weights $w_i$ were set to $p_{T}$ in the TPC region and the nMIP values from the EPD with a minimum threshold of 0.3 and maximum of 2.0. In addition, all weights for tracks or hits with $\eta<-1.045$ have a negative sign while all with $\eta>-1.045$ are positive to account for the the fact that $v_{1}$ is odd in rapidity. Each event used in the analysis was required to have a minimum of five hits in the inner EPD region, nine hits in the outer EPD region, and five tracks in the TPC region to ensure that each subevent had an adequate number of particles to reconstruct $\Psi_1$.

Before calculating $R_{31}$ and flow, we perform recentering followed by Fourier shifting corrections on all event plane distributions to remove biases from non-uniform detector acceptance \cite{Poskanzer}. 
After these corrections, all event plane distributions are isotropic from $-\pi$ to $\pi$. The values calculated for the event plane resolution in each centrality are shown in Fig. \ref{fig:resolutions} (see the Supplemental Material \cite{supplemental} for a comparison to $R_{11}$). These specific values were calculated considering systematic effects and are described in the next section.

\begin{figure}[!h]
    \centering
    \includegraphics[scale=0.32]{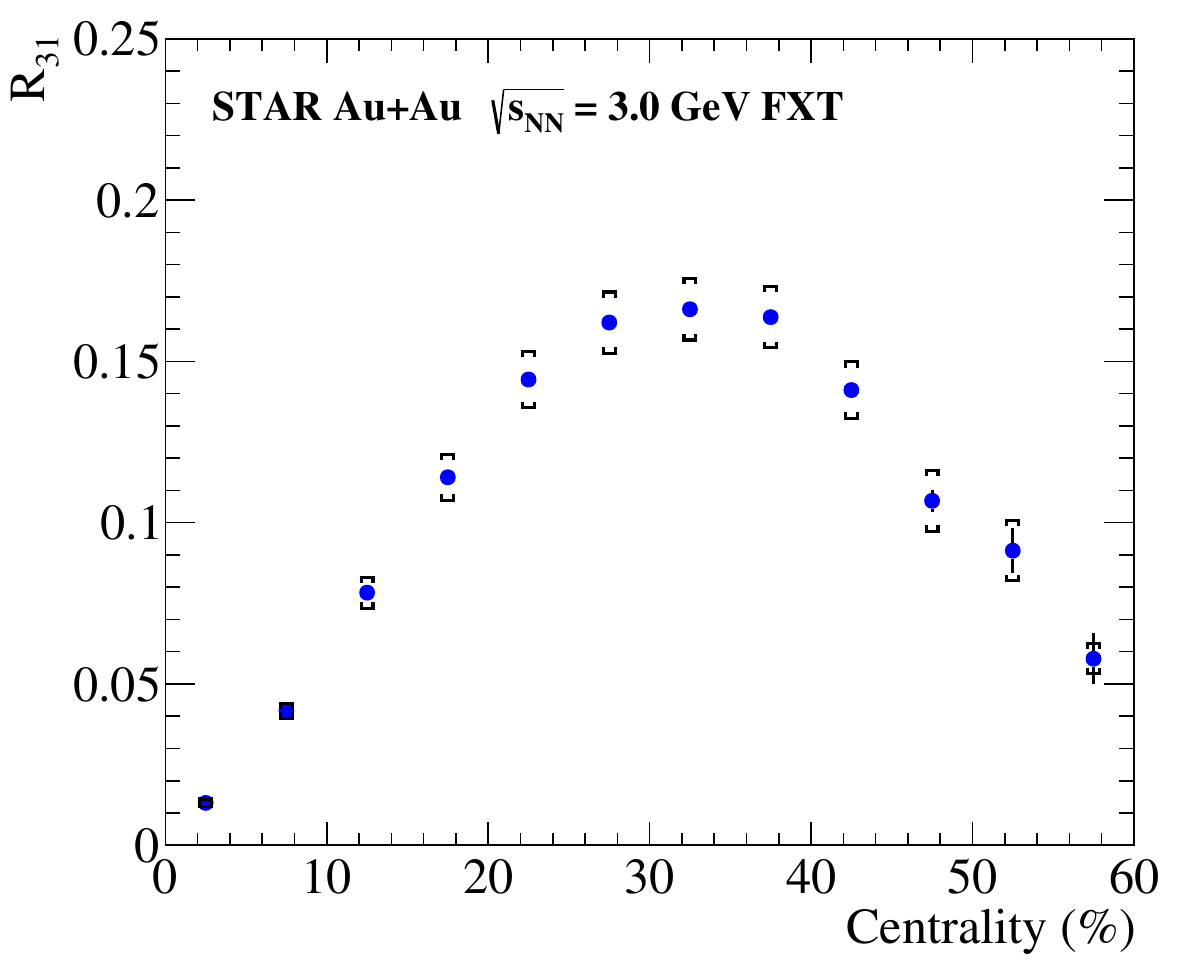}
    \caption{Event plane resolution for $v_3\{\Psi_1\}$ as a function of centrality from $\sqrt{s_{NN}}=3$ GeV Au+Au collisions at STAR. The points are the average of the resolutions from the three configurations discussed in Sec. \ref{sec:systematics}, and the systematic uncertainties are the maximum difference between the configurations and the average (taken as a symmetric uncertainty in the opposite direction as well). Vertical lines are statistical uncertainties and open brackets are systematic uncertainties.}
    \label{fig:resolutions}
\end{figure}

\subsection{\label{sec:systematics}Systematic uncertainties}
We varied all track quality and particle identification cuts
by 20\%, and again by 30\%, to estimate the systematic uncertainties contributed by each cut and denoted these as $\sigma_{\mathrm{sys,}i}$, where $i$ would be any of the cuts mentioned in Secs. \ref{sec:selection} and \ref{sec:particleID}. The $\sigma_{\mathrm{sys,}i}$ was calculated as the standard deviation from all five measurements of $v_3\{\Psi_1\}$.
The final systematic uncertainty for any one $v_3\{\Psi_1\}$ value is calculated as $\sqrt{\sum_{i} \sigma^2_{\mathrm{sys,}i}}$, where the summation only includes the cuts with variations deemed significant for that measurement \cite{Barlow}. However, the variation for the event plane resolution does not change the amount of data analyzed (like a variation on the vertex would, for example). Therefore, the variation on the event plane resolution was always included in the summation for systematic uncertainties in every $v_3\{\Psi_1\}$ measurement. In this way, the systematic uncertainty for any measurement is always greater than or equal to the systematic uncertainty contribution from the event plane resolution variation.

The maximum values of the resolution, $R_{31}$, were obtained by utilizing all three subevents described in Sec. \ref{sec:EPA}.
To estimate systematic errors, two variations were implemented to modify the size of region $B$ in the EPD to introduce an $\eta$ gap between regions $A$ and $B$. This was done to eliminate potential momentum conservation effects and assess other systematic factors involved in the calculation of $R_{31}$. One variation involved removing one ring of tiles from EPD B closest to EPD A, creating a gap between the two. This reduced EPD B to $-3.1<\eta<-2.5$. The other variation removed two rings from EPD B in the same way, changing its coverage to $-2.8<\eta<-2.5$.
By employing the two variations for the event plane from region $B$, in addition to the initial configuration, an envelope of $R_{31}$ values was established. The average value $\langle R_{31} \rangle$ from all three configurations was adopted as the resolution, and the maximum deviation from this average among the three variations was taken as the systematic error for $R_{31}$ (see Fig. \ref{fig:resolutions}). To propagate the uncertainties in $R_{31}$ to variations in flow, the resolutions are increased/decreased by the total uncertainty (statistical plus systematic in quadrature) and flow is calculated.

Table \ref{table:systematicContributions} shows the contribution of each systematic source as a percentage of $v_3\{\Psi_1\}$, averaged across all measurements, in three wide centrality regions. All contributions related to track quality cuts were combined in quadrature and listed in the first row; and the same was done for event selection cuts shown in the second row. The proton $\langle dE/dx \rangle$ contribution in 0-10\% centrality is around 54\% simply due to the very small magnitude of the proton $v_3\{\Psi_1\}$ in this centrality range (particularly in $v_3\{\Psi_1\}$ vs rapidity plots, which were not calculated for pions and kaons).

\begin{table}[!htb]
    \begin{center}
        \caption{Average contribution of each varied cut to systematic uncertainties in $v_{3}\{\Psi_{1}\}$ as a percentage of the $v_{3}\{\Psi_{1}\}$ value for three centrality ranges.}
        \label{table:systematicContributions}
        \begin{tblr}{ hlines,vlines,hspan=even,cells={c,m} }
            \SetCell[c=1]{c} \textbf{Systematic source} 
                & \SetCell[c=3]{c} \textbf{Uncertainties in percent} \\ 
            Centrality interval & 0-10\% & 10-40\% & 40-60\% \\ \hline
            Track quality & 13.5 & 3.0 & 3.9 \\
            Event quality & 2.8 & 0.3 & 0.7 \\
            $\pi$ $\langle dE/dx \rangle$ & 6.0 & 2.8 & 3.4 \\
            $K$ $\langle dE/dx \rangle$ & 5.7 & 4.1 & 11.3 \\
            Proton $\langle dE/dx \rangle$ & 53.8 & 3.1 & 3.2 \\
            TOF $m^2_{\pi}$ & 3.1 & 1.0 & 1.4 \\
            TOF $m^2_{K}$ & 13.1 & 13.5 & 7.4 \\
            Event plane resolution & 7.7 & 4.9 & 9.9 \\
        \end{tblr}
    \end{center}
\end{table}

\section{\label{sec:level4}Results}
\begin{figure}[!htb]
    \centering
    \includegraphics[scale=0.32]{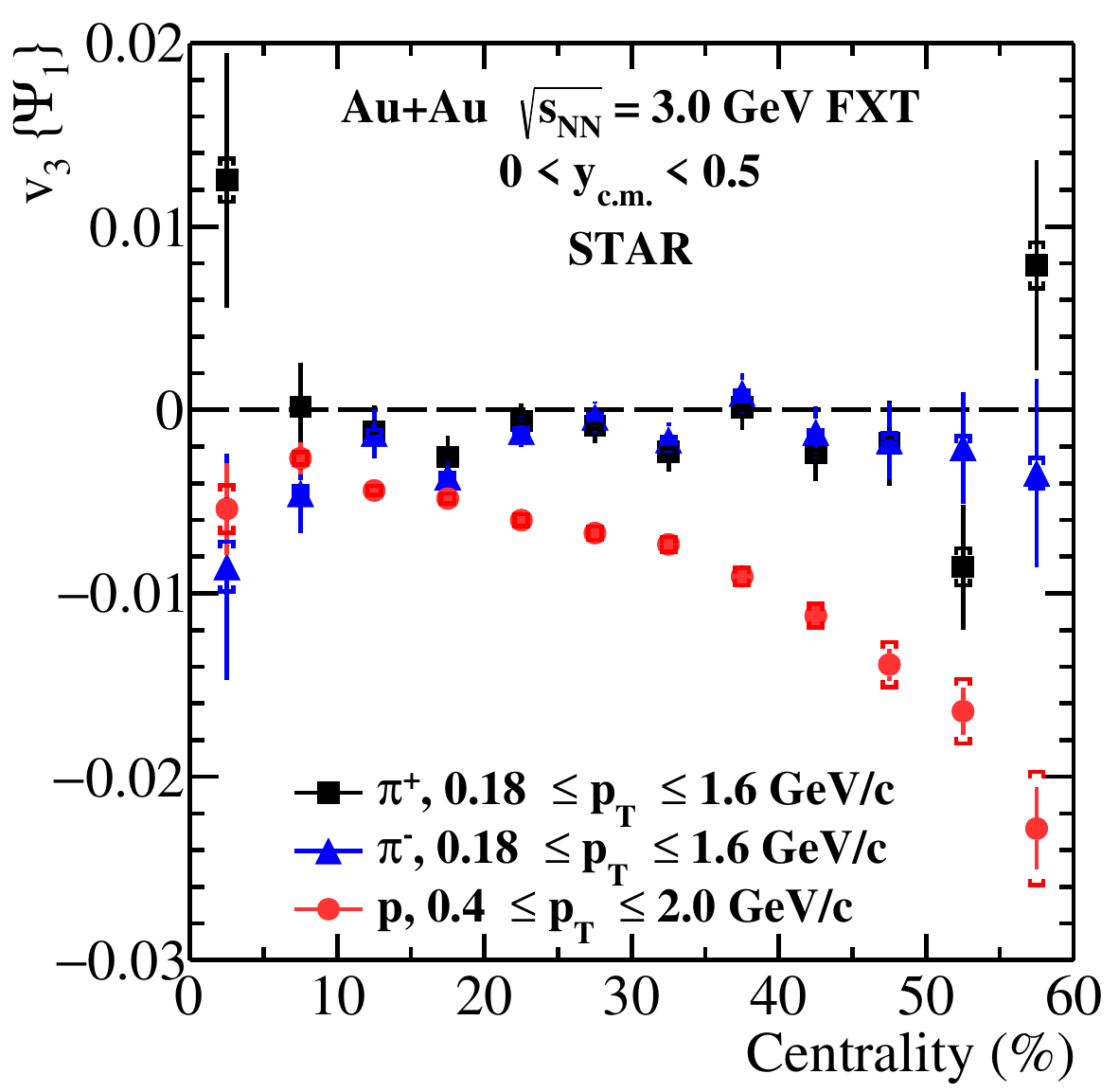}
    \caption{$v_3\{\Psi_1\}$ vs. centrality for $\pi^+$, $\pi^-$, and protons using the event plane method. Protons show a clear negative $v_3\{\Psi_1\}$ while pions remain near zero. Statistical uncertainties are shown as lines while systematic uncertainties are denoted by open square brackets.}
    \label{fig:piAndPrCentrality}
\end{figure}

\begin{figure}[!htb]
    \centering
    \includegraphics[scale=0.32]{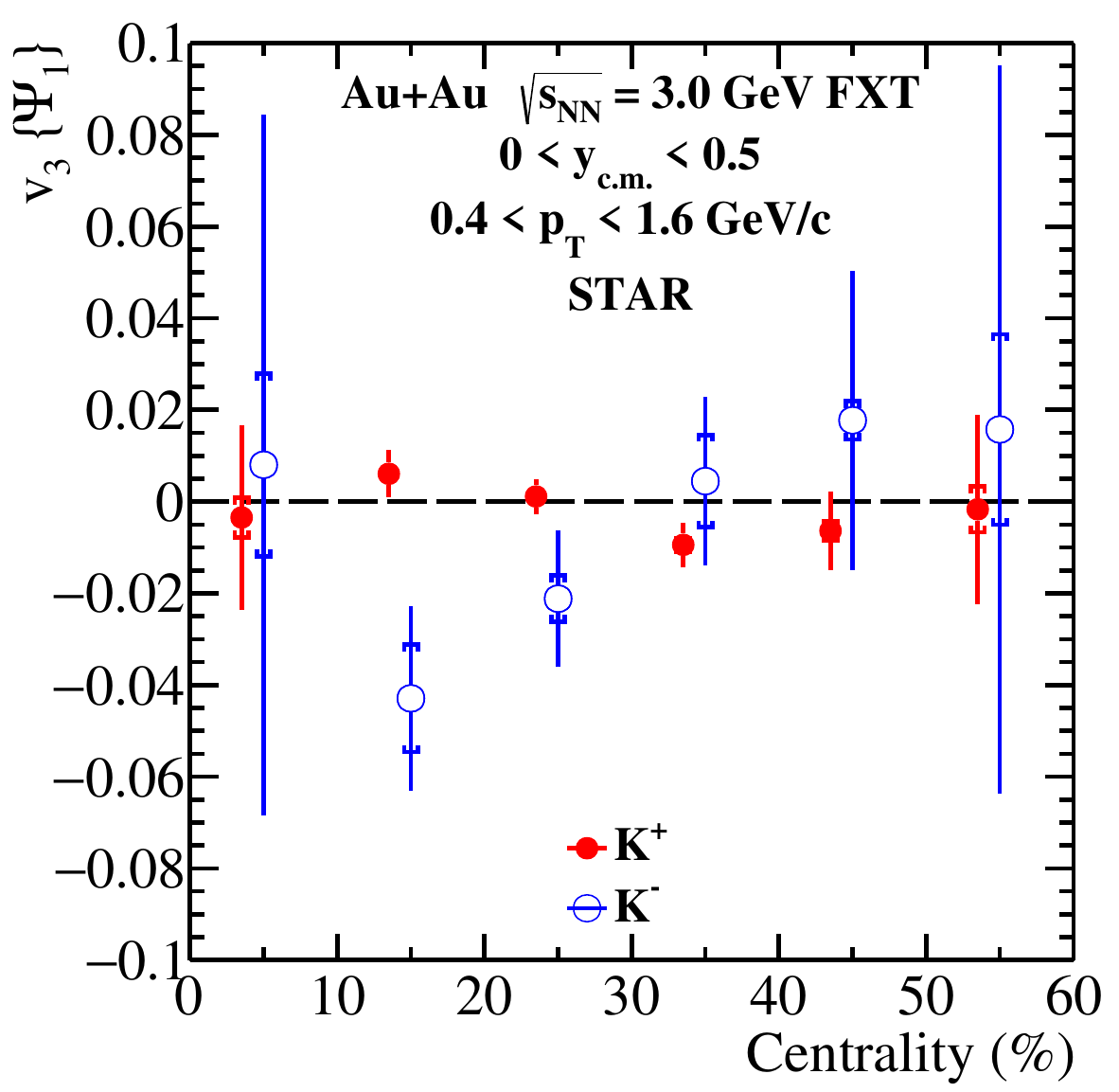}
    \caption{$v_3\{\Psi_1\}$ vs. centrality for $K^+$ and $K^-$ using the event plane method. The values for $K^+$ are slightly shifted horizontally for visual clarity. More statistics are required to fully understand if this signal is present for kaons. Statistical uncertainties are shown as lines while systematic uncertainties are shown as open square brackets.}
    \label{fig:kaCentrality}
\end{figure}

The $v_3\{\Psi_1\}$ values for $\pi^{\pm}$, $K^{\pm}$, and protons were calculated as a function of centrality and these results are presented in Figs. \ref{fig:piAndPrCentrality} and \ref{fig:kaCentrality}. Figure \ref{fig:piAndPrCentrality} specifically shows that, while the values for pions are only slightly negative, the protons exhibit a significant negative $v_3\{\Psi_1\}$ signal that increases with centrality. Figure \ref{fig:kaCentrality} suggests that $v_3\{\Psi_1\}$ for $K^+$ may also be consistent with zero, while the available statistics for $K^-$ do not allow for a definitive conclusion. To test the significance of the $K^-$ signal we performed a fit using a constant value at zero. This produced $\chi^2/\mathrm{NDF}=6.9/5=1.38$, hence we can not rule out the possibility that this signal is consistent with zero.

\begin{figure}[!htb]
    \centering
    \includegraphics[scale=0.32]{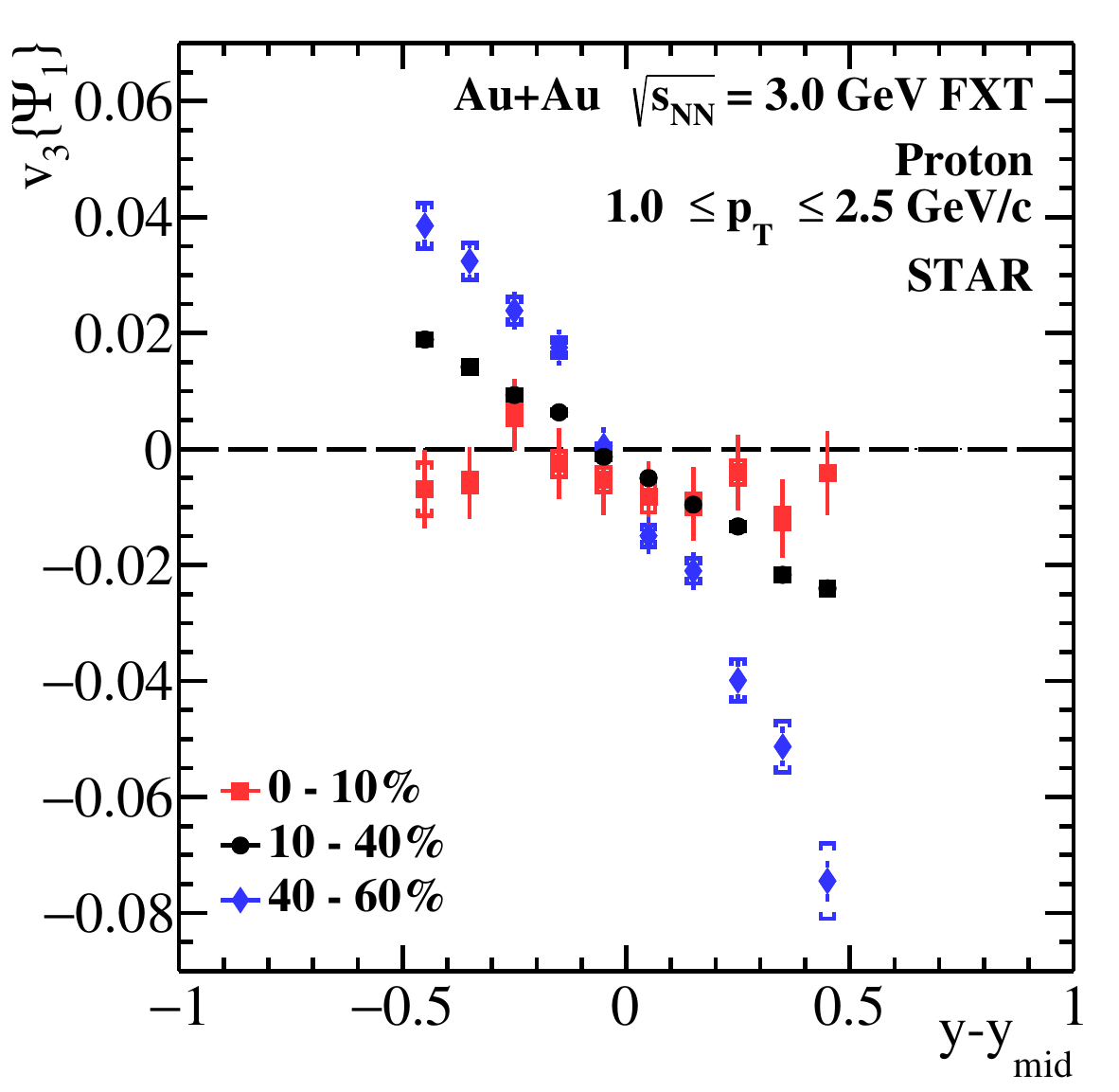}
    \caption{$v_3\{\Psi_1\}$ vs. rapidity for protons in three large centrality bins from a symmetric acceptance across midrapidity. Protons exhibit an increasingly negative slope going towards more peripheral collisions. Statistical uncertainties are shown as lines while systematic uncertainties are denoted by open square brackets.}
    \label{fig:prSymmetric}
\end{figure}

Figure \ref{fig:prSymmetric} displays $v_3\{\Psi_1\}$ values as a function of rapidity in a symmetric acceptance region indicated by the dashed black box in Fig. \ref{fig:normalAcceptance} where the $p_{T}$ is required to be above 1.0 GeV/$c$ in order to have an acceptance below midrapidity. Here we see that the $v_3\{\Psi_1\}$ signal is essentially rapidity-odd, however this is not exact. The points do not trace straight through (0, 0), and in particular for 40-60\% centrality, the points in the backward region reach about $-0.08$ while those in the forward only reach about $0.04$. We measure the event plane in the negative rapidity side of the collision only, so there may be effects such as event plane rapidity decorrelations, and our measurement of $v_3\{\Psi_1\}$ may not be exactly rapidity-odd. However, we will speak of $v_3\{\Psi_1\}$ as rapidity-odd since the signal's source is not primarily fluctuations that are even in rapidity, and since our measurements do closely resemble a rapidity-odd behavior.
Previous STAR results at this energy \cite{STAR:2021yiu, STAR:2021ozh} show a positive $(dv_1/dy)_{y=0}$ slope for protons. In contrast, Fig. \ref{fig:prSymmetric} shows a clearly negative $(dv_3\{\Psi_1\}/dy)|_{y=0}$.

\begin{figure}[!htb]
    \centering
    \includegraphics[scale=0.32]{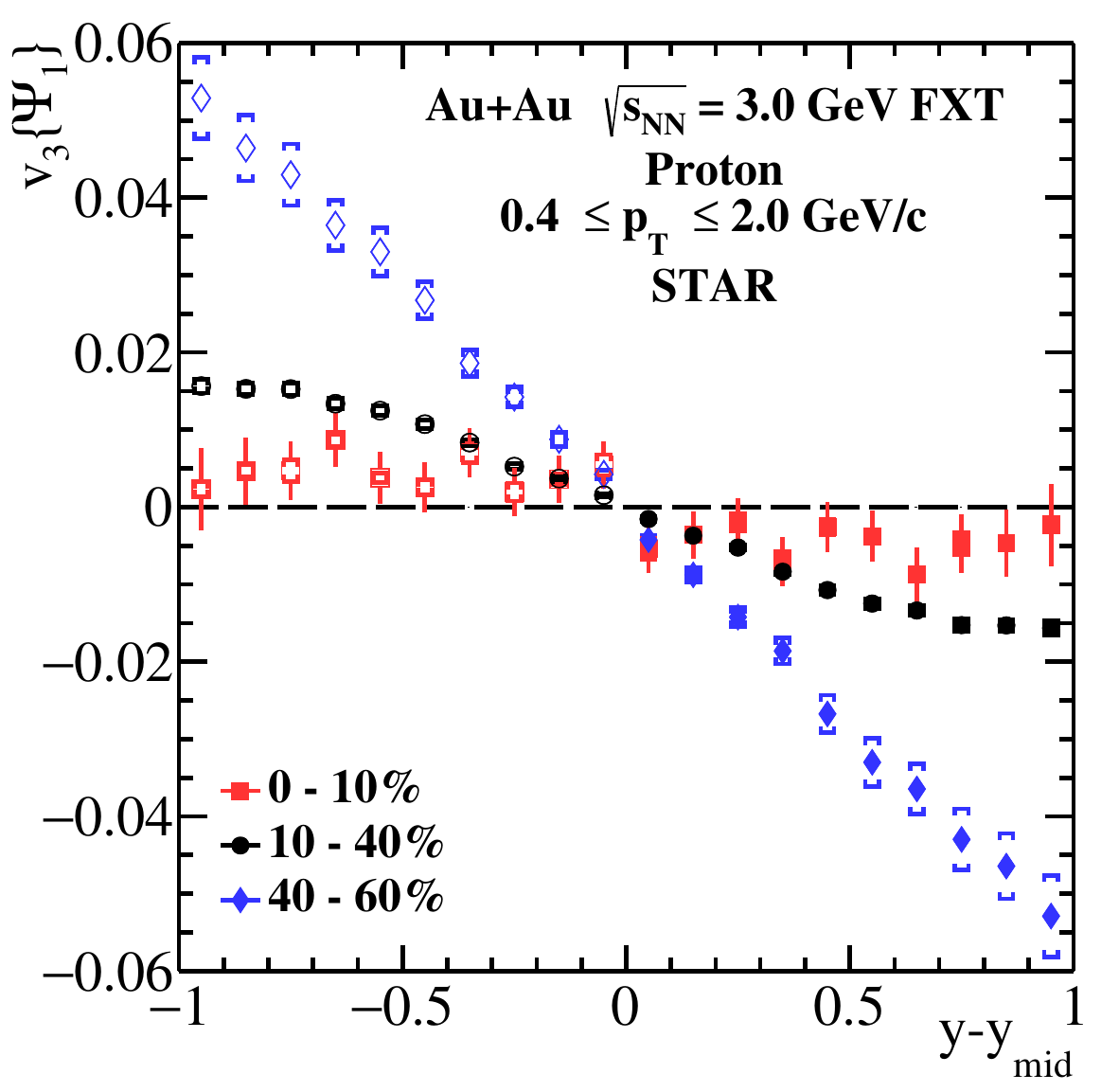}
    \caption{$v_3\{\Psi_1\}$ vs. rapidity for protons in three large centrality bins from only the backward region (solid markers) along with mirrored points across midrapidity (open markers). Note that the $p_{\mathrm{T}}$ acceptance extended to a lower limit than in Fig. \ref{fig:prSymmetric}. Statistical uncertainties are represented as lines while systematic uncertainties are denoted by open square brackets.}
    \label{fig:prNotSymmetric}
\end{figure}

In Fig. \ref{fig:prNotSymmetric}, we additionally present $v_3\{\Psi_1\}$ as a function of rapidity in the backward region $0 < y_{\mathrm{c.m.}} < 1$, extending the $p_{T}$ range down to 0.4 GeV/$c$ and we have mirrored the measured points and represented them as open circles. By fitting the function $y=ax+bx^3$, where $y$ is $v_3\{\Psi_1\}$, $x$ is $y-y_{\mathrm{mid}}$, and $a$ is $d v_3/dy|_{y=0}$, to the 10-40\% centrality data using only the measured points, we obtain a slope of $d v_3/dy|_{y=0}=-0.025 \pm 0.001$ (stat).

To estimate the systematic error on the slope, we assume that the percentage difference between slope measurements for the positive and negative rapidity sides of Fig. \ref{fig:prSymmetric} would be similar to the percentage difference for the positive and negative sides (if measurable) of Fig. \ref{fig:prNotSymmetric}. We utilize both sides of Fig. \ref{fig:prSymmetric} to calculate $\Delta a/(\langle a \rangle \sqrt{12}) = 0.13$, where we have assumed the two slopes as a continuous uniform distribution to get a standard deviation of $\Delta a/\sqrt{12}$. The systematic error for the slope in Fig. \ref{fig:prNotSymmetric} is then determined as $|-0.025| \times 0.13 \approx 0.003$. Therefore, the final measurement of the slope is $d v_3/dy|_{y=0}=-0.025 \pm 0.001$ (stat) $\pm 0.003$ (sys).

\begin{figure}[!htb]
    \centering
    \includegraphics[scale=0.32]{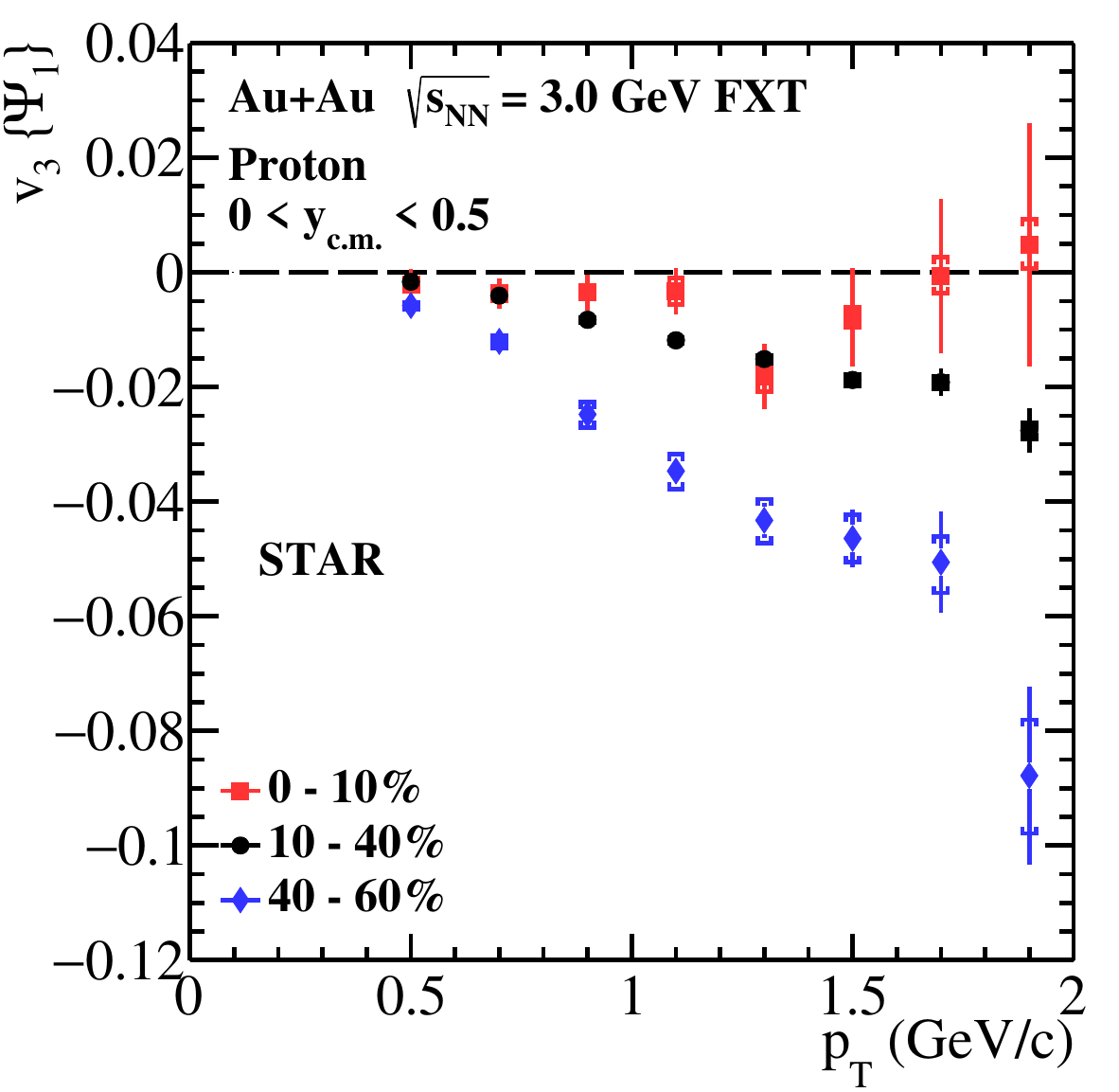}
    \caption{$v_3\{\Psi_1\}$ vs. $p_{\mathrm{T}}$ for protons in three large centrality bins. $v_3\{\Psi_1\}$ becomes increasingly negative as $p_{\mathrm{T}}$ and centrality increase. Statistical uncertainties are shown as lines while systematic uncertainties are denoted by open square brackets.}
    \label{fig:prPt}
\end{figure}

In Fig. \ref{fig:prPt} we present $v_3\{\Psi_1\}$ values from protons vs. $p_{\mathrm{T}}$, showing that the magnitude increases with increasing $p_{\mathrm{T}}$.  

Figures \ref{fig:prSymmetric} - \ref{fig:prPt} show that the proton $v_3\{\Psi_1\}$ values become increasingly negative towards more peripheral collisions, consistent with Fig. \ref{fig:piAndPrCentrality}, with the effect being strongest at the largest $p_{\mathrm{T}}$ and rapidity.

\section{\label{sec:level5}Discussion }
This section will first compare models to data. Subsequently, we provide a more comprehensive discussion of the models, specifically focusing on how the initial geometry and potentials are manifest with particular emphasis on protons. Lastly, a brief discussion of pions and kaons and a comparison to the previous HADES result will follow.

\subsection{Model comparisons}
The results have been compared with several models: AMPT, RQMD, SMASH and JAM \cite{Lin:2004en,Bass:1998ca,Weil:2016zrk,Nara:2019qfd}. For the sake of illustration, we will focus on the JAM and SMASH models. URQMD includes similar physics to the JAM model, while AMPT does not include mean field potentials, which, as is discussed below, provides a crucial element to describe the data and understanding the physics. 

Each model includes a cascade mode which treats baryons and mesons as individual particles modeling their interactions as if they were colliding billiard balls. The radius of each particle is determined by the cross section and secondary particles are formed according to known reactions. There is no long range interaction. 

\begin{figure}
    \centering
    \includegraphics[scale=0.43]{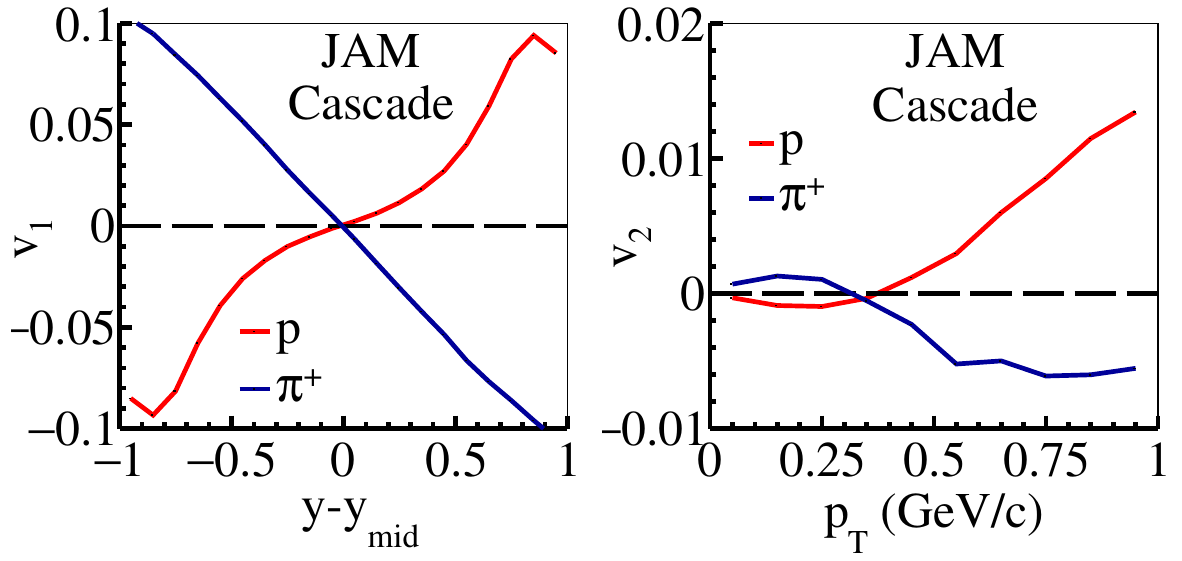}
    \caption{$v_1$ vs. rapidity (left) and $v_2$ vs. $p_{\mathrm{T}}$ (right) for protons and $\pi^+$ in 3 GeV center of mass ``minimum bias'' Au+Au collisions as given by the JAM model in cascade mode.}
    \label{fig:v1_v2_jam_cascade}
\end{figure}

\begin{figure}
    \centering
    \includegraphics[scale=.32]{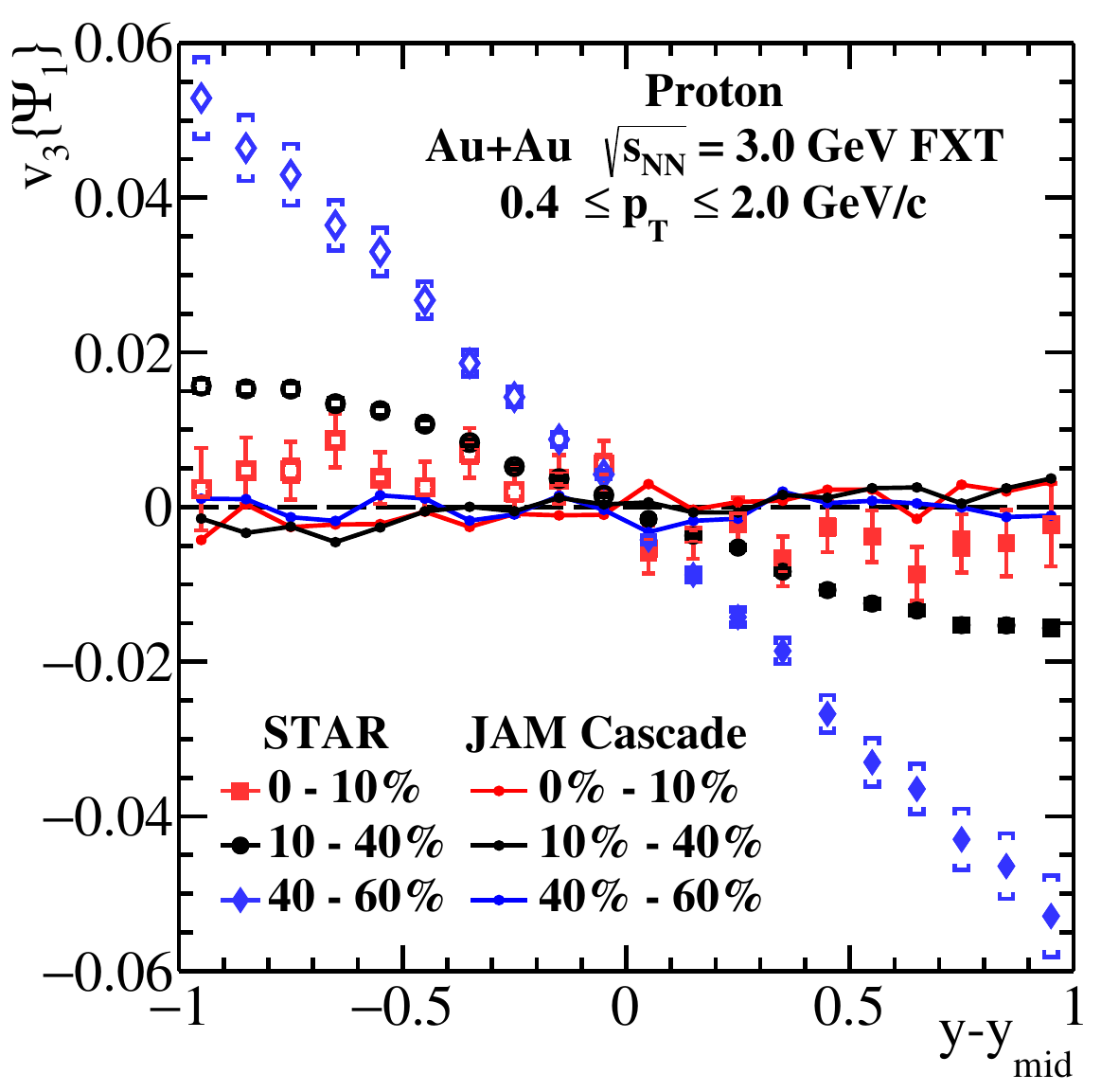}
    \includegraphics[scale=.32]{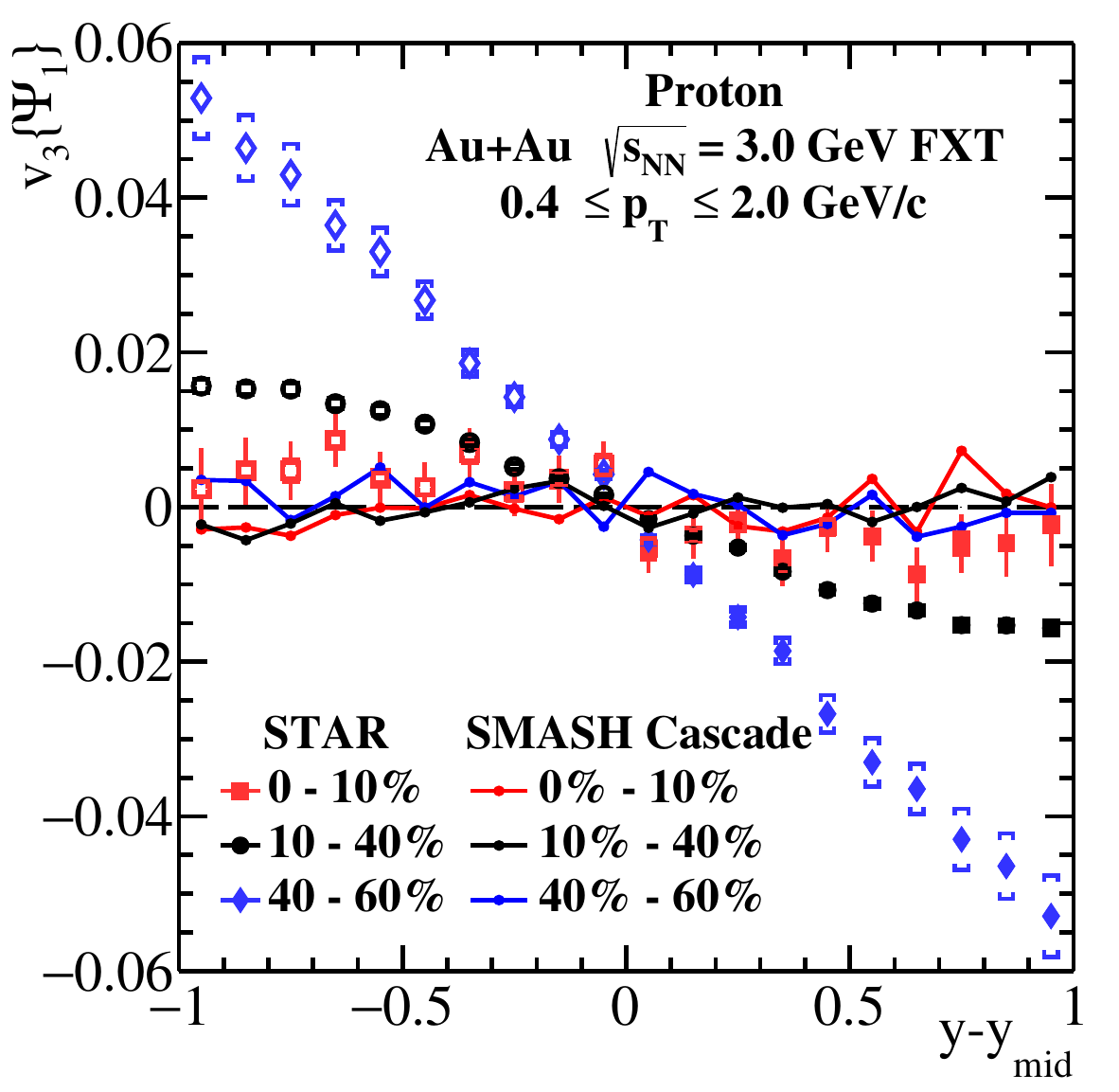}
    \caption{$v_3\{\Psi_1\}$ for protons in several centralities in the JAM model (top) and SMASH model (bottom) as compared to the data. Cuts used in the model are identical to the data, where centrality is defined by cuts in impact parameter as described in the text.}
    \label{fig:v3_smash_jam_cascade}
\end{figure}

At high center-of-mass energy, above $\sqrt{s_{NN}} \approx  30$ GeV, the timescale of the collision is very short relative to the formation time (typically about 1 fm/$c$) and geometric effects are essentially two-dimensional in the $x$-$y$ plane. However, at lower energies, the passing time of the two nuclei in a Au+Au collision is long compared to the formation time; at $\sqrt{s_{NN}}=3$ GeV it is $\approx10$ fm/$c$. This is the energy regime in which baryon stopping becomes dominant \cite{E917:2000spt}. Such a scenario leads to effects described as ``bounce off'' leading to a finite $v_1$ and ``squeeze out" leading to a finite $v_2$ \cite{Bass:1998ca}. Both $v_1$ and $v_2$ can be generated using models in their cascade mode as illustrated for the JAM model in Fig. \ref{fig:v1_v2_jam_cascade}, although the magnitudes do not accurately describe the data. In contrast, the observed $v_3\{\Psi_1\}$ cannot be generated by a cascade model as shown in Fig. \ref{fig:v3_smash_jam_cascade}. In the cascade mode, virtually no $v_3\{\Psi_1\}$ is developed in the models. As cascade models are incapable of generating $v_3\{\Psi_1\}$, the detection of $v_3\{\Psi_1\}$ necessitates an alternate driving force which can be provided by a nuclear potential as described in section \ref{sec:meanfield}. 

\begin{figure*}
    \centering
    \includegraphics[scale=.29]{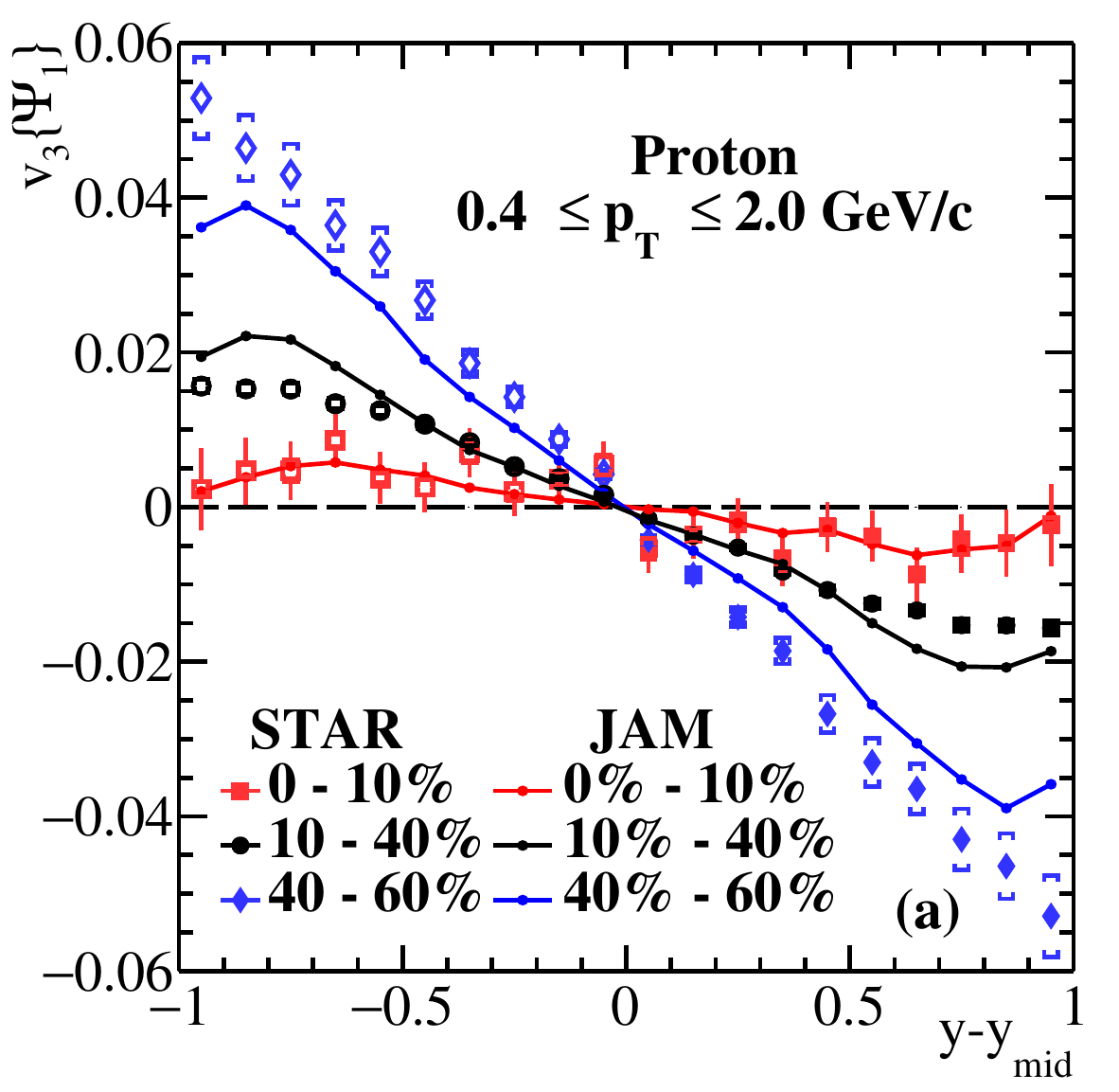}
    \includegraphics[scale=.29]{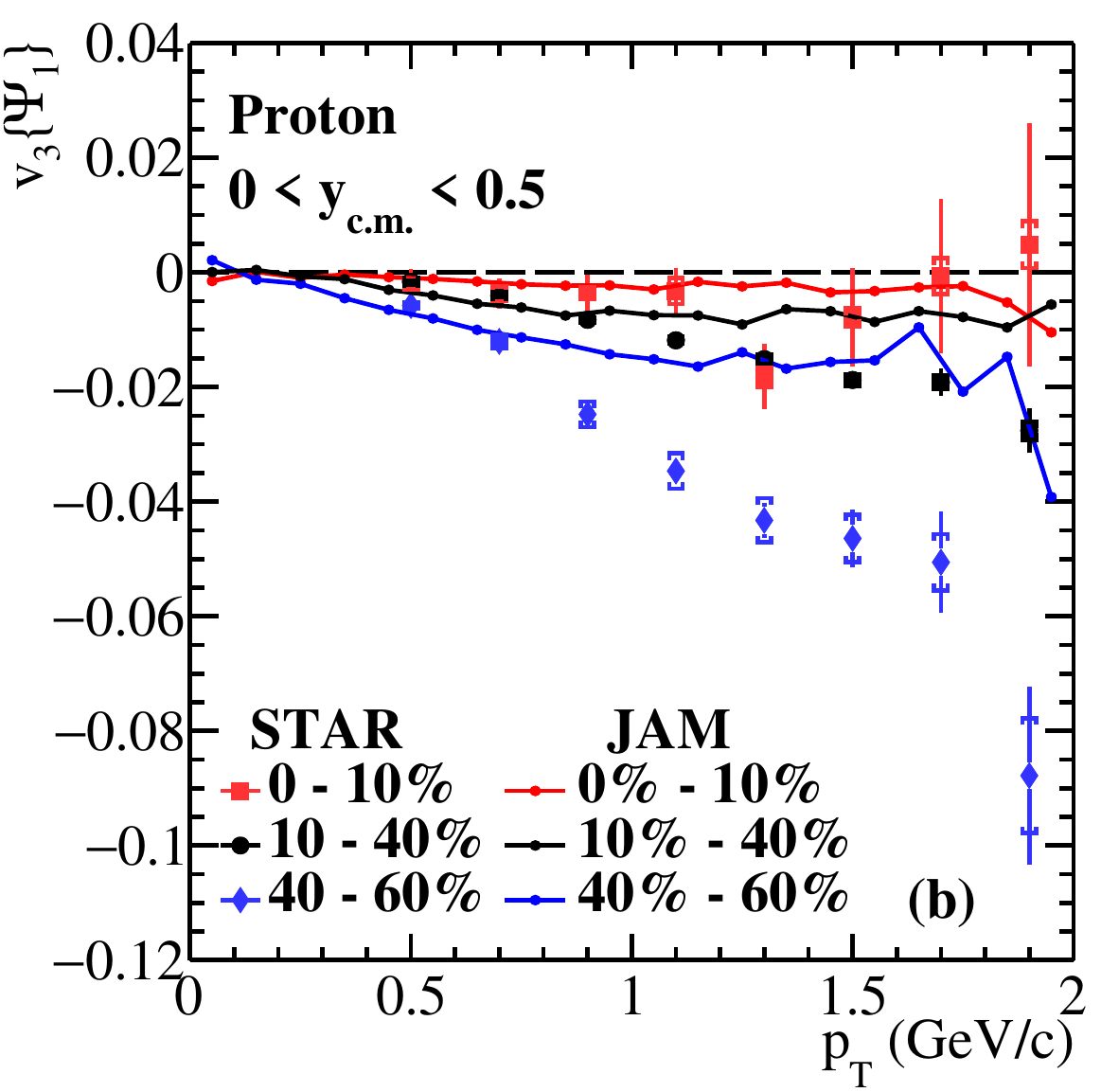}
    \includegraphics[scale=.29]{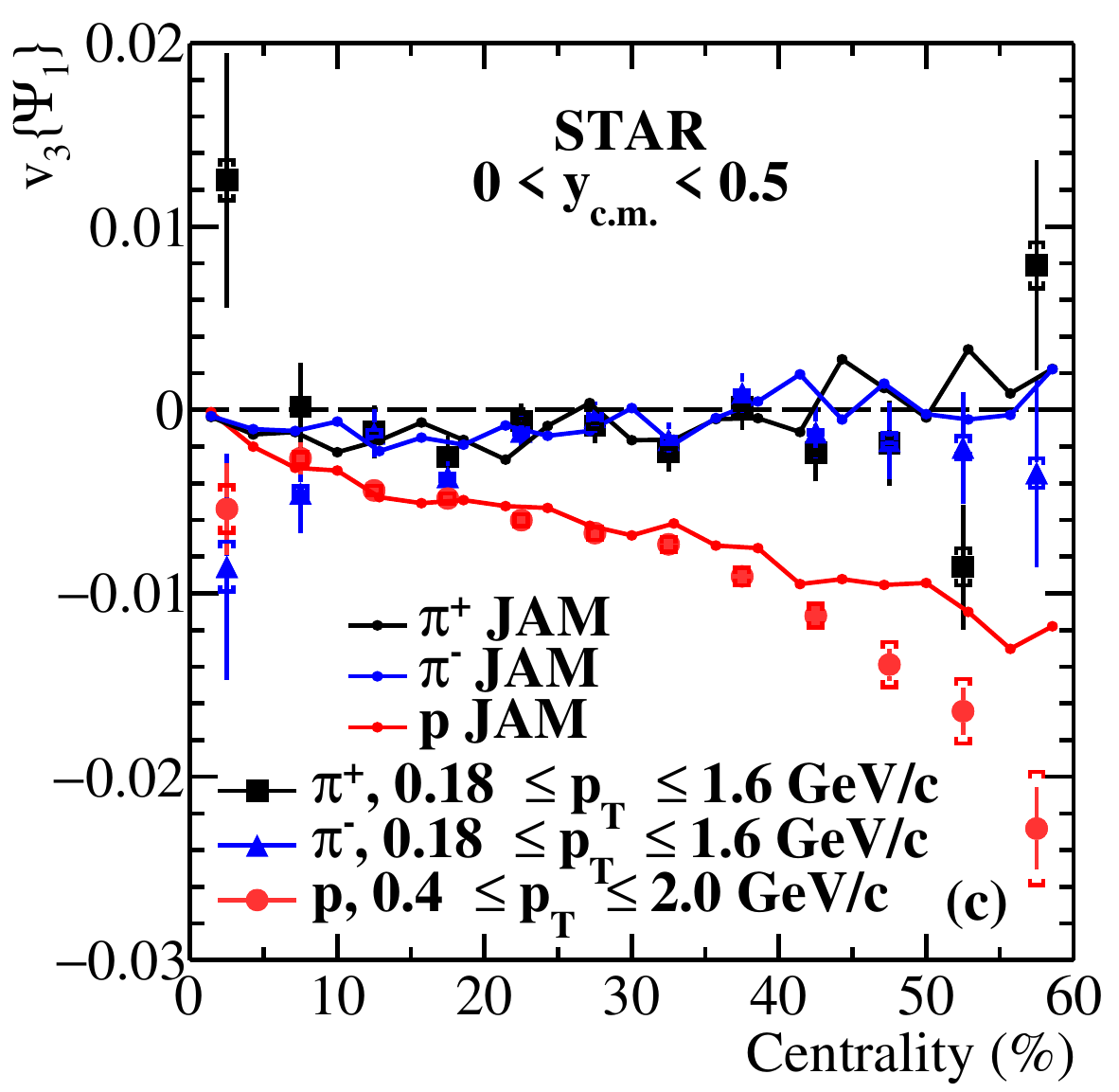}
    \includegraphics[scale=.29]{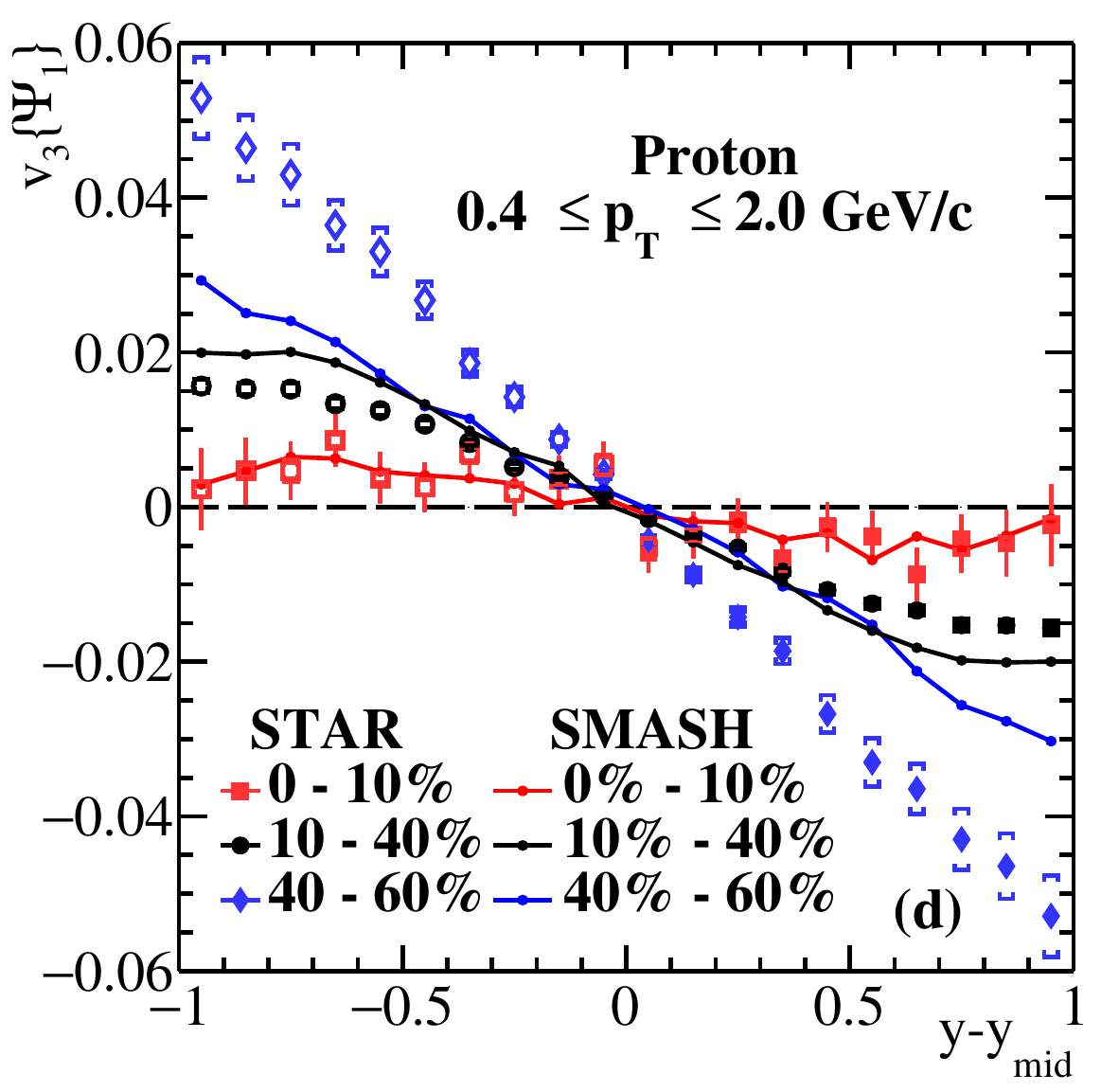}
    \includegraphics[scale=.29]{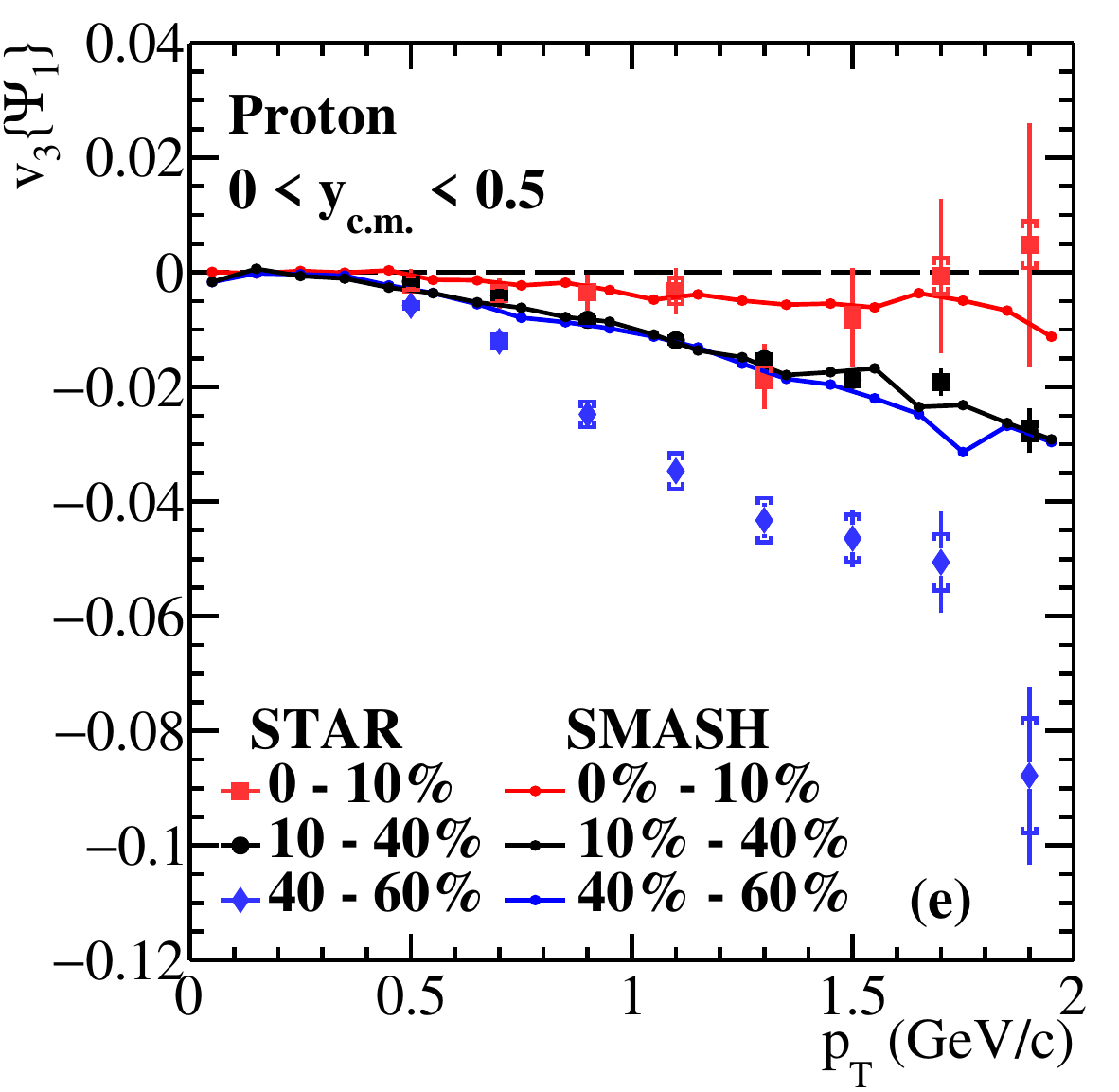}
    \includegraphics[scale=.29]{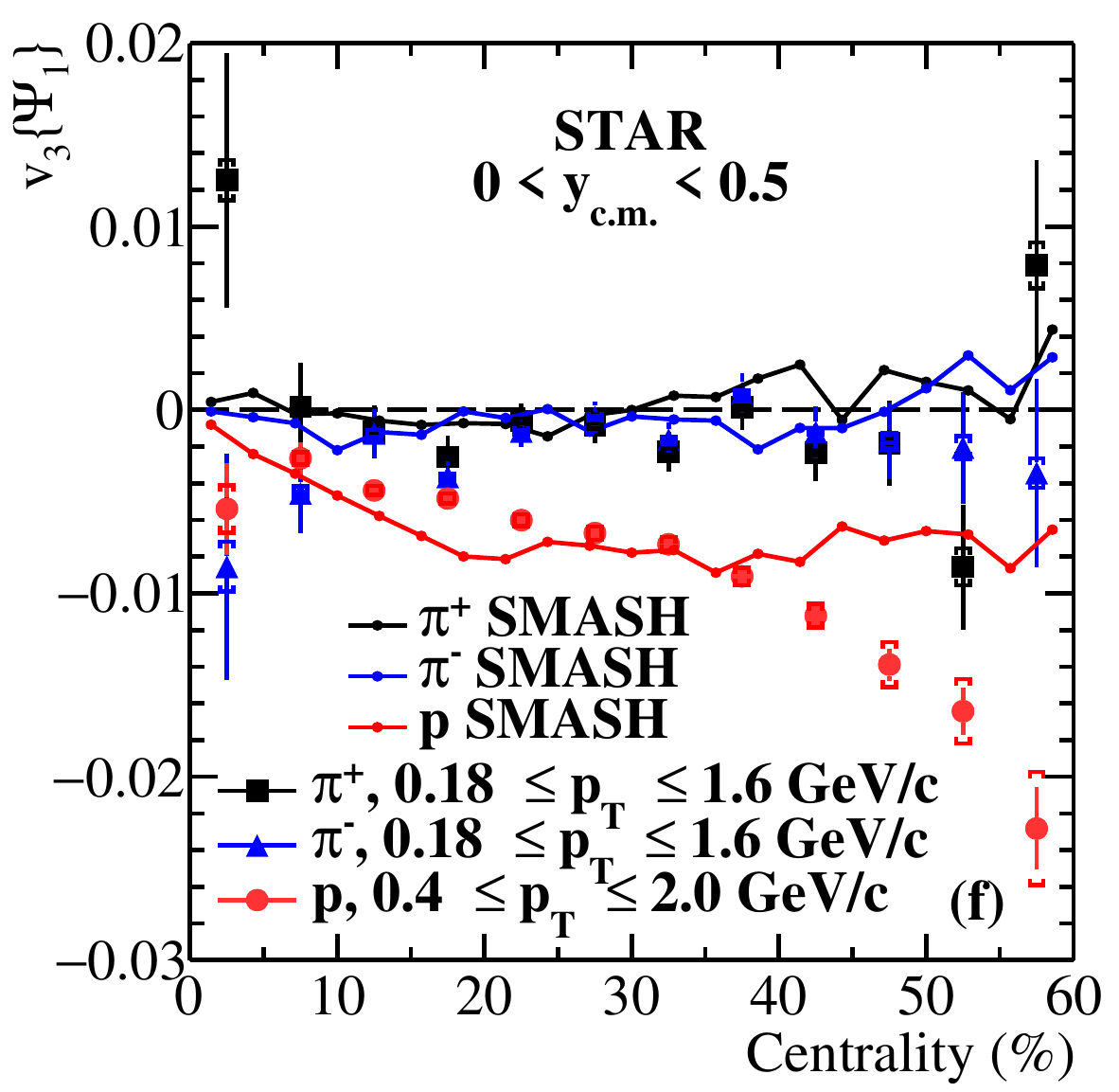}
    \includegraphics[scale=.29]{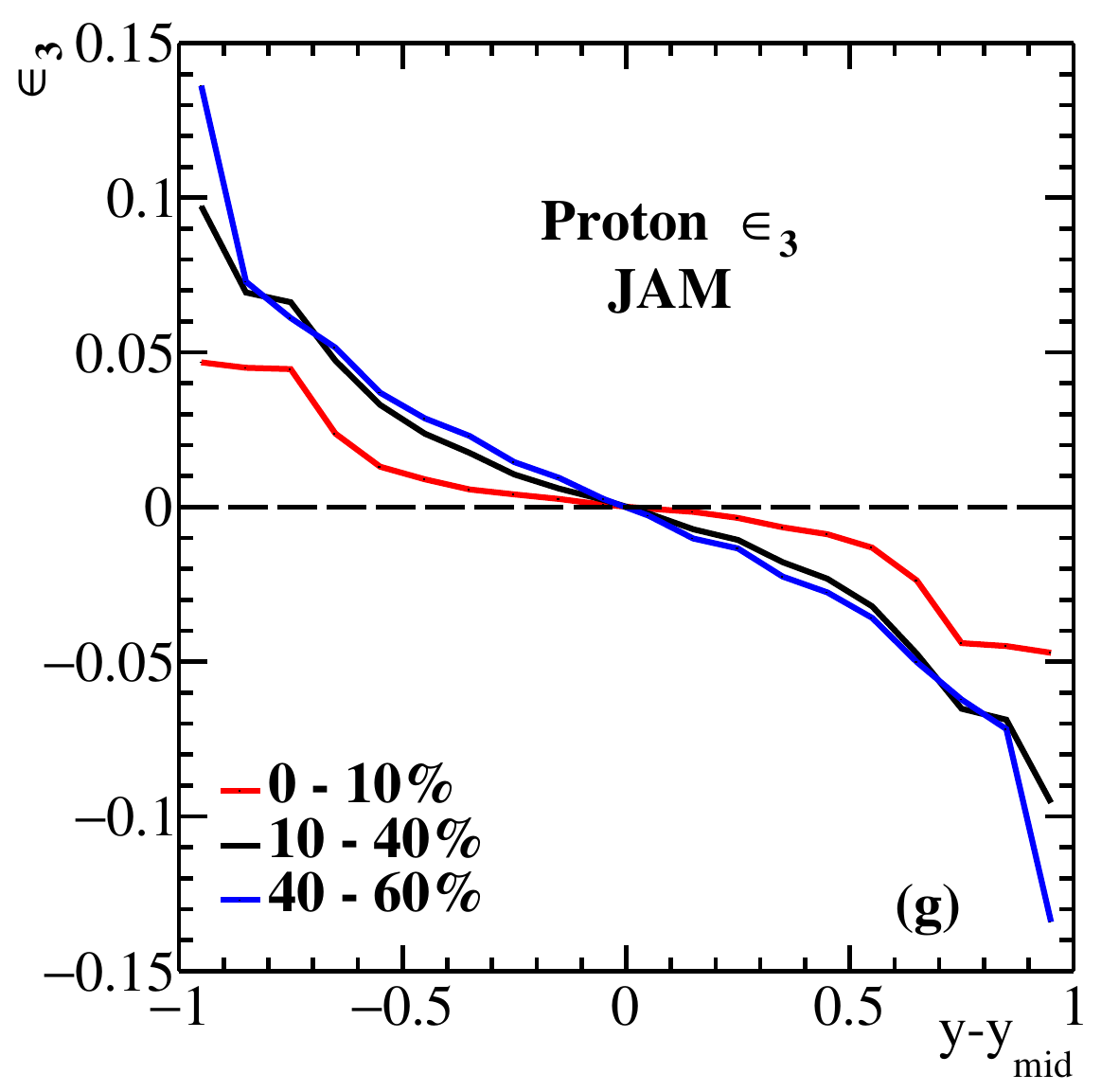}
    \includegraphics[scale=.29]{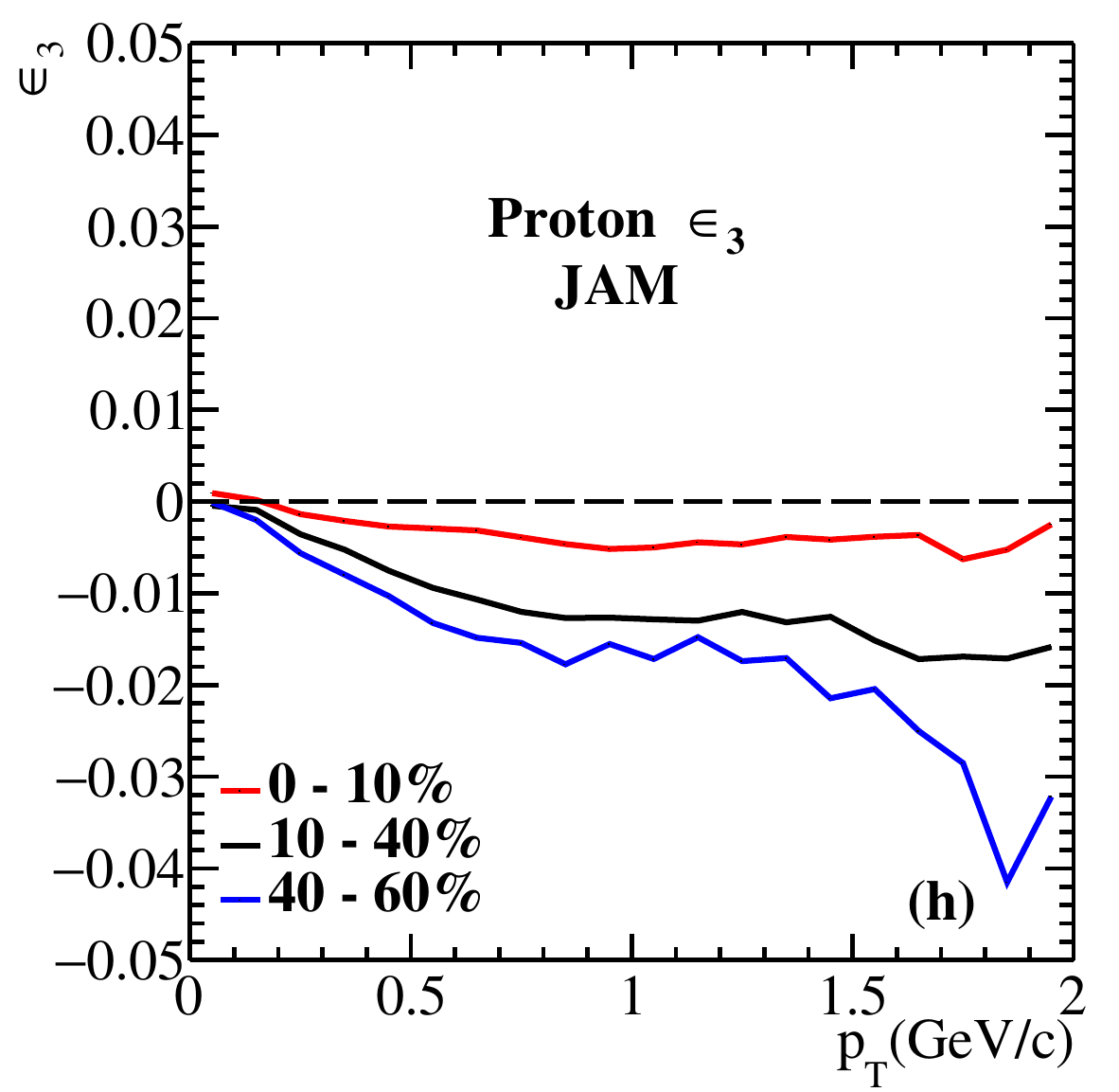}
    \includegraphics[scale=.29]{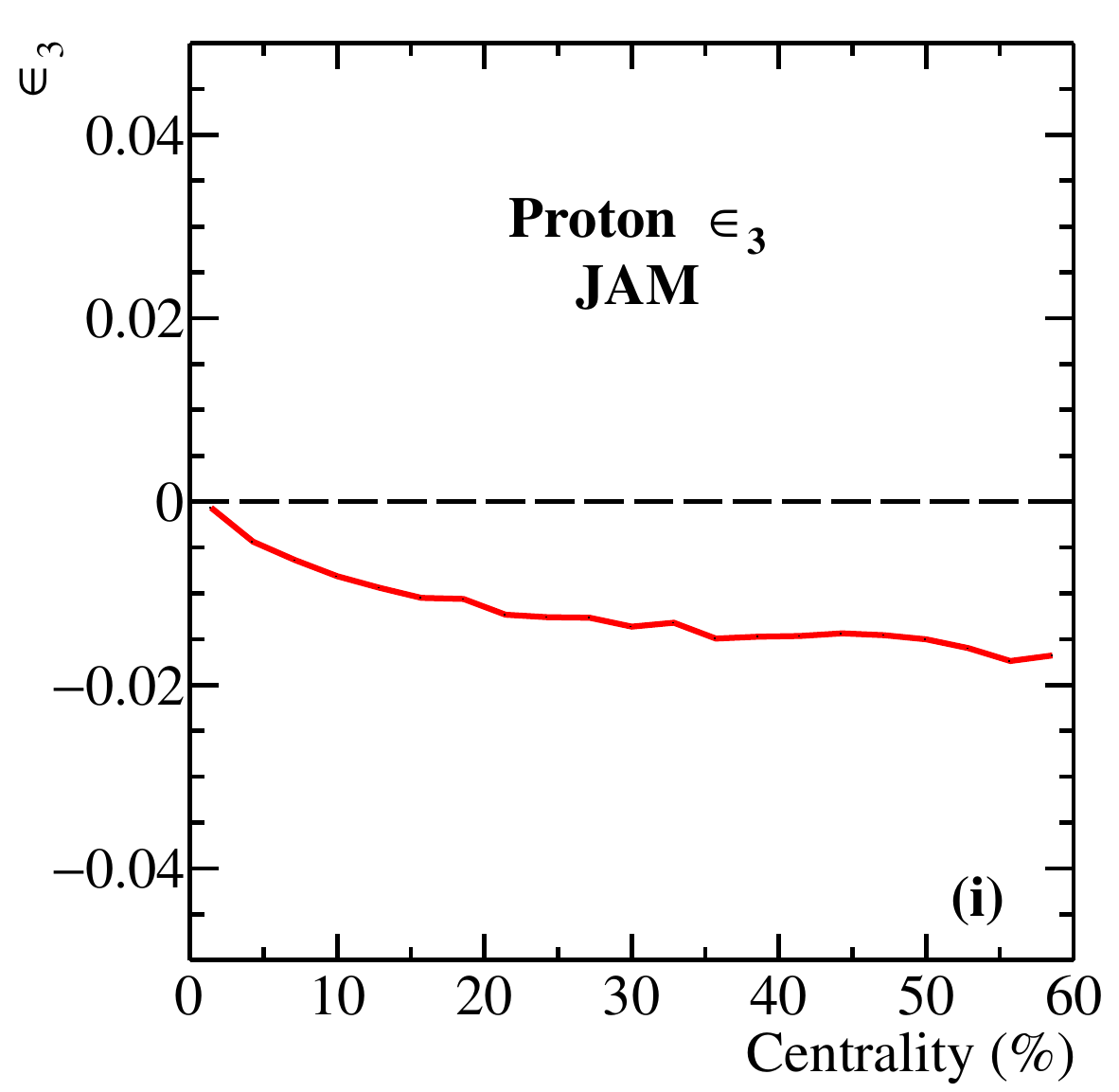}
    \caption{The top two rows show fits of JAM [top row, (a)-(c)] and SMASH [second row, (d)-(f)] to the $v_3\{\Psi_1\}$ data vs. rapidity (first column), $p_{\mathrm{T}}$ (second column) and centrality (third column). The fits to distributions vs. rapidity and $p_{\mathrm{T}}$ (a), (b), (d), (e) are fits to protons, for three centrality bins. Hollow points in (a), (d) are reflected around the mid-rapidity as explained in the text. Fits to centrality (c), (f) show protons, $\pi^+$, and $\pi^-$. The bottom row (g)-(i) depicts $\epsilon_3$ in the JAM simulation for protons at $t=20$ fm/$c$ vs  rapidity, $p_{\mathrm{T}}$, and centrality. }
    \label{fig:v3_fits}
\end{figure*}

Figures \ref{fig:v3_fits}(a-c) and \ref{fig:v3_fits}(d-f) show comparisons of the data with JAM and SMASH simulations, respectively, where potentials have been included in the models. JAM $v_3\{\Psi_1\}$ and SMASH $v_3\{\Psi_1\}$ 
values are shown vs. rapidity, $p_{\mathrm{T}}$, and centrality. For JAM and SMASH vs. centrality, we show values for $\pi^+$, $\pi^-$ and protons. (The triangularity, $\epsilon_3$, will be discussed later.) 

Note that centrality for the models uses cuts on the impact parameter $b$, where we assume that the nucleus is a spherical ball with radius 6.64 fm.
The introduction of potentials reproduces the trends of the $v_3\{\Psi_1\}$ observed in the data as a function of rapidity, $p_{\mathrm{T}}$, and centrality. Both models appear to have a weaker response for peripheral collisions than the data, although JAM is slightly better; this is reflected in the rapidity and $p_{\mathrm{T}}$ distributions as well as the centrality distributions for peripheral events [Figs. \ref{fig:v3_fits}(e-f)]. JAM has a slightly weaker response, i.e. smaller $v_3\{\Psi_1\}$, than SMASH at higher $p_{\mathrm{T}}$ for mid-central collisions [compare Figs. \ref{fig:v3_fits}(b) and \ref{fig:v3_fits}(e)]. 

\subsection{Role of collision geometry}
\begin{figure}
    \centering
    \includegraphics[scale=0.21]{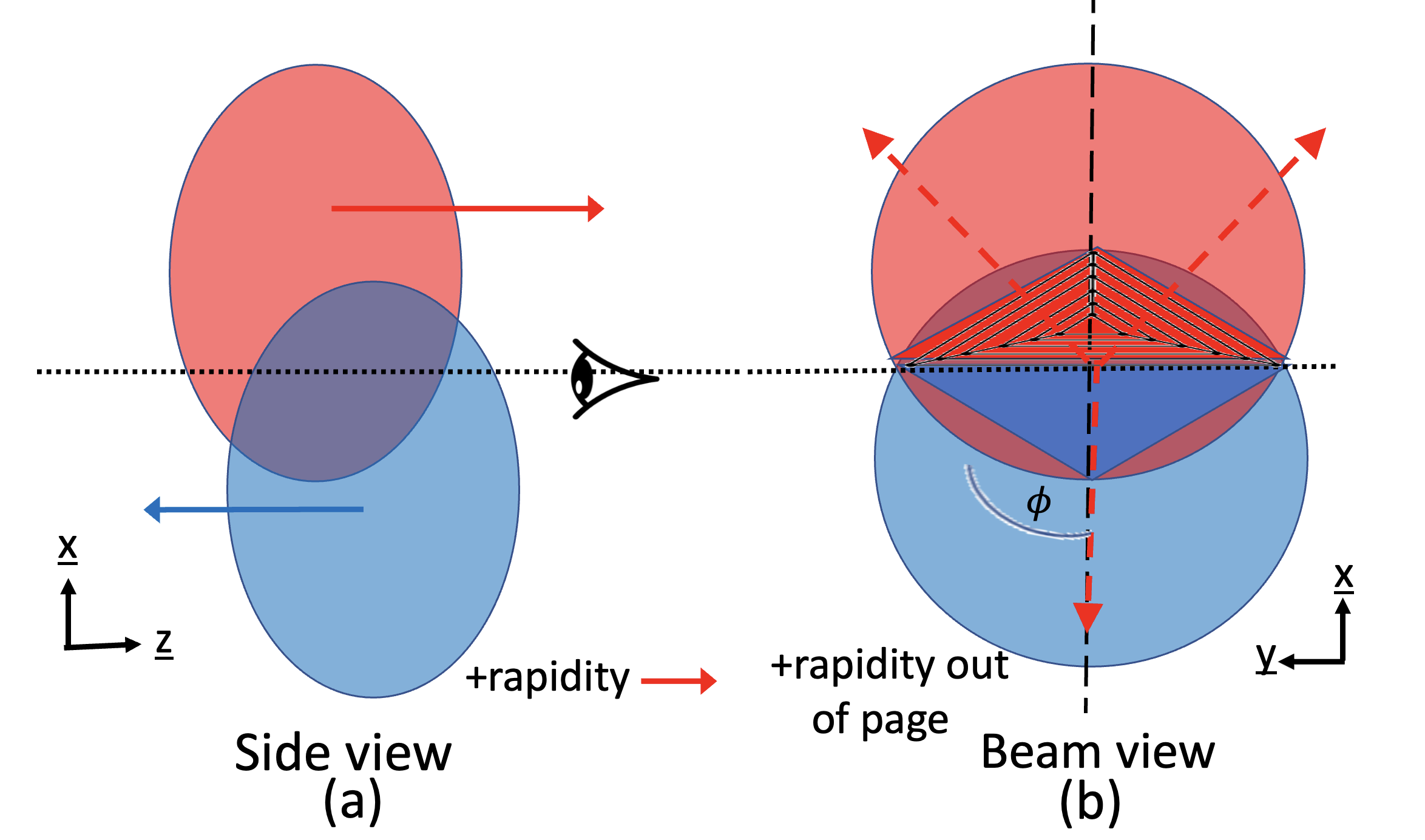}
    \caption{Illustration of the triangular geometry of participants as discussed in the text. \newline (a) Side view. Red nucleus is moving towards positive rapidity, blue nucleus to negative rapidity. The collision bifurcates the participants as indicated by the dotted line, where the nucleus from one half must penetrate through a thicker portion of the other nucleus, thereby experiencing more stopping and creating half-moon, triangular shapes. The red triangle [more easily visualized in (b)] moves towards positive rapidity, blue triangle towards negative rapidity. The eye indicates the beam view as shown in (b). \newline(b) Beam view. Positive rapidity is out of the page.
    The familiar almond shape formed by the participants resulting in $v_2$, primarily at midrapidity, is evident. However, at higher rapidity, the nucleons forming the half-moon shape with nonzero triangular component from the red nucleus penetrates through the thin tip of the blue nucleus, while the remainder of the nucleons from the red nucleus which participate in the collision are stopped by the blue nucleus. The triangular shape emerges from the usual almond-shaped collision region, half of which is obstructed by the opposing nucleus. The lines in the red triangle portray the pressure gradients along the directions of the red arrows, resulting in a negative $v_3\{\Psi_1\}$.}
    \label{fig:triangle_geometry_cartoon}
\end{figure}

The non-central collision of two large nuclei at low energies where stopping is strong, can result in a triangular geometry at rapidities away from $y_{\mathrm{c.m.}}=0$. The stopping is a result of a combination of the nucleon-nucleon cross section, the collision energy which sets the time-scales, and the nuclear thickness. The origin of the triangular shape primarily stems from the uneven stopping by one nucleus onto the other in non-central events, with stronger stopping on one side for a specific rapidity region, as depicted in Fig. \ref{fig:triangle_geometry_cartoon}(a). The red triangle's lines in Fig. \ref{fig:triangle_geometry_cartoon}(b) portray the gradients leading to pressure in the red arrows' direction, yielding a negative $v_3\{\Psi_1\}$. 

\begin{figure}
    \centering
    \includegraphics[scale=0.17]{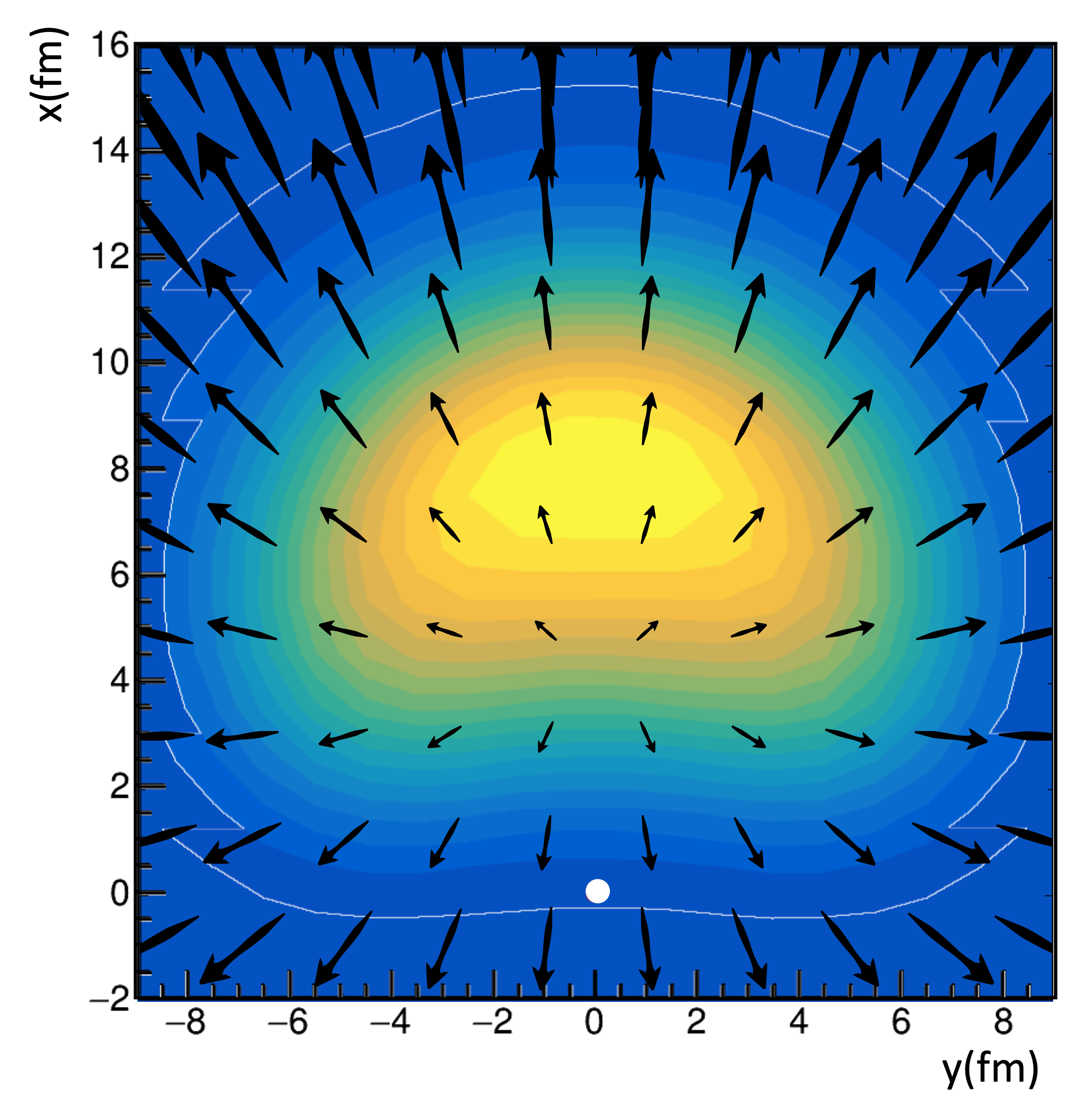}
    \caption{
    Illustration of the $x$ vs. $y$ position of protons from JAM at $t=50$ fm/$c$ for particles with rapidity $0.6 <y< 0.85$ and $0<p_{\mathrm{T}}<2$ GeV/$c$ (avoiding spectators) illustrating the half-moon, ``triangular'' shape. Note that the high density region is centered at $x\approx7$ fm. The arrows depict the vector field obtained by dividing the collision region into cells. The average momentum in each cell is superimposed on the original plot. The length of each arrow represents the magnitude of the average momentum in each cell. The white dot indicates the $x=0$, $y=0$ position.}
    \label{fig:momentum_vectors}
\end{figure}

In order to study the relevant geometry in simulations where we can examine particles in configuration space, we select participant protons by requiring them to have a rapidity between $0.6 <y<0.85$ and $0 < p_{\mathrm{T}} < 2$ GeV/$c$. We then use the models to plot the x and y positions of the protons in this region at $t=50$ fm/$c$ from the beginning of the simulation, when the flow has fully developed, but the triangular geometry is still visible even though some expansion of the medium has already occurred. Figure \ref{fig:momentum_vectors} illustrates the spatial configuration of all such protons in 40-60\% centrality collisions.  We ensure the rapidity to be away from beam rapidity ($y=1.05$) to select participants, although at low center-of-mass energy the separation of participants and spectators is not as apparent as at higher energies since there is time for the particles in the participant region to be scattered into the spectator region, and vice-versa. The half-moon shape, yielding the spacial distribution with a third harmonic (i.e. the triangular shape) necessary to seed $v_3\{\Psi_1\}$, is clearly visible. 

In analogy to the eccentricity (or ellipticity) we define triangularity as follows \cite{Teaney:2010vd}: 
\begin{equation}
\epsilon_3=-\frac{\langle r^2 \cos(3(\phi-\Psi_r))\rangle}{\langle r^2 \rangle}.
\end{equation}
Here, $\phi$ represents the angle between the particle and the reaction plane. The reaction plane angle is set to $\Psi_r = 0$ in both SMASH and JAM for all events.
In these calculations, the origin was reset to the center of the distribution in the particular rapidity slice of interest.
Figures \ref{fig:v3_fits}(g)-\ref{fig:v3_fits}(i) display the results for $\epsilon_3$ using the JAM model at $t=20$ fm/$c$, considering particles sorted by rapidity or $p_{\mathrm{T}}$. The time $t=20$ fm/$c$ was chosen to allow spectators to distinguish themselves from the participants; the flow, which would dilute the spacial shape, was not yet fully developed.

The trends in triangularity, $\epsilon_3$, are clearly reflected in the data, indicating that the initial geometry, as quantified by the triangularity, provides the necessary shape to seed the $v_3\{\Psi_1\}$ observed in the data. The observed flow is a function of both the shape and the potential which we discuss next.

\subsection{\label{sec:meanfield}Role of mean field potentials}
\begin{figure}
    \centering
    \includegraphics[trim={1cm 0cm 0cm 0cm},scale=.4]{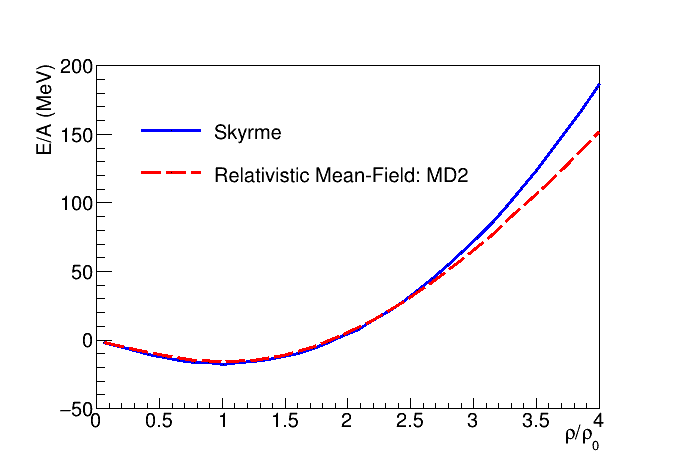}
    \caption{Total energy per nucleon for the potential used in the models.  The relatively hard Skyrme potential with $K=380$ MeV used in the SMASH model is shown as the blue-solid line. The relativistic mean field potential with parameter set MD2 used in the JAM model is shown as the dashed red line \cite{Nara:2020ztb}}.
    \label{fig:potential}
\end{figure}

Mean field potentials have been used to describe collisions with nuclear targets at low energies. Several early models, successful in describing heavy-ion collisions, incorporated hadronic potentials \cite{Serot:1997xg, Buss:2011mx,Blaettel:1993uz}. 
The potentials in these models are typically a function of the baryon density as shown in Fig. \ref{fig:potential}. In this case, the potential is smaller towards smaller baryon density for $\rho/\rho_0 > 1$, resulting in a repulsive force in that region.
Among the potentials were nonrelativistic Skyrme potentials and relativistic mean field potentials. The latter incorporated exchange particles such as the $\sigma$ (scalar-isoscalar) and $\omega$ (vector-isoscalar) mesons. As center-of-mass energies increase, relativistic effects become more prominent. Some potentials allow for the sensitivity to phase transitions; however to accurately account for the transition to the quark-gluon plasma, a transition to a hydrodynamic simulation is probably needed. 
As mentioned in the introduction, previous studies have shown that for $\sqrt{s_{NN}} \lesssim 4$ GeV, mean field effects are important to explain observed directed and elliptic flow \cite{Buss:2011mx,Bass:1998ca, Bleicher:1999xi,FOPI:2004bfz}. This also holds true in our case.

To generate $v_3\{\Psi_1\}$, the introduction of a baryon density dependent potential was necessary. In the SMASH model, primarily aimed at low energy collisions below $\sqrt{s_{NN}} = 3$ GeV, a Skyrme+symmetry potential was used. Fermi motion and Pauli blocking were incorporated. The potential in SMASH is taken as 
\begin{equation}
U=A(\rho/\rho_0)+B(\rho/\rho_0)^{\tau} \pm 2 S_{\mathrm{pot}} \frac{\rho_{I_3}}{\rho_0},
\end{equation}
where $\rho$ is the baryon density and $\rho_{I_3}$ is the baryon isospin density of the relative isospin projection $I_3/I$. $\rho_0$ = 0.1681/fm$^3$ is the nuclear ground state density. Parameters for the Skyrme potential are $A=-124.0$ MeV, $B=71$ MeV, and $\tau = 2$. For the symmetry potential, $S_{pot}=18$ MeV and the positive and negative signs refer to neutrons and protons, respectively. This model also reproduces an incompressibility of $K=380$ MeV. These values are taken from values used by URQMD, which gave reasonable fits to preliminary HADES data on $v_1$, $v_2$ and $v_3\{\Psi_1\}$ \cite{Hillmann:2018nmd}.

We employed a relativistic mean field in the JAM1 model (RQMD.RMF)\footnote[3]{While the term RQMD is often used as the name of the code developed by Sorge, Stoecker and Greiner \cite{VONKEITZ1991601}, RQMD within the term RQMD.RMF refers to the underlying theoretical $N$-body model in JAM \cite{Nara:2019qfd}.}. 
The potential invokes a relativistic mean field theory incorporating $\sigma$- and $\omega$-meson-baryon interactions and momentum-dependent potentials as described in \cite{Nara:2020ztb}. The parameter set MD2, described in the reference, has the same incompressibility as the SMASH model we used ($K=380$ MeV). The parameter set yields results that are consistent with numerous data sets on sidewards flow $\langle p_x \rangle$ from midcentral Au+Au collisions from E895 and E877 at $\sqrt{s_{NN}}=2.7\textrm{--}4.86$ GeV. STAR and NA49 $v_1$ from midcentral Au+Au at $\sqrt{s_{NN}}<8.87$ GeV are also consistent with JAM using the MD2 parameter set. It is also consistent with the recent $\sqrt{s_{NN}} = 3$ GeV proton directed and elliptic flow results \cite{STAR:2021yiu}.
However, above $\sqrt{s_{NN}}=8.87$ GeV, an additional attractive orbit is required, consistent with a softening of the equation of state (EOS). Figure \ref{fig:potential} depicts the energy per nucleon vs. $\rho / \rho_0$ used in the two models in this work. 
Note that higher baryon density regions are to the right and the force will be towards lower baryon density regions. This will naturally produce a pressure away from regions of high baryon density, typically outwards in a collision.  

To visualize the motion of the medium, the density plot shown in Fig. \ref{fig:momentum_vectors} is divided into cells and the average momentum is computed for each cell. The resulting vector field is superimposed on the original plot clearly showing the outward motion of the medium. Both the momentum vector and the density of the medium must be considered to obtain the flow. The white dot in the center indicates the origin in the $x$-$y$ plane. The snapshot is taken at a time for $t=50$ fm/$c$ when the flow has had time to develop fully but the spacial configuration of the nucleons is still visible. 

We now have identified the two essential ingredients required to generate $v_3\{\Psi_1\}$: (1) the initial condition supplied by the triangular shape resulting from a combination of the centrality and stopping, and (2) the force on the medium supplied by the potential. 

\subsection{Pions and kaons}
Figures \ref{fig:piAndPrCentrality} and \ref{fig:kaCentrality} indicate that $v_3\{\Psi_1\}$ for $\pi^+$, $\pi^-$, and $K^+$ are essentially zero within $\approx$ 0.5\%. $v_3\{\Psi_1\}$ for $K^-$ may be negative, but that is at most a 1$\sigma$ effect. The potentials used in the models only act on baryons. Any effect from these potentials on pions or kaons, such as on $v_1$, stem from secondary interactions resulting from the multiple scattering off baryons or from decays \cite{Agnieszka}. The models do not generate any $v_3\{\Psi_1\}$ for pions or kaons, hence $v_3\{\Psi_1\}$ appears to be a uniquely sensitive probe for a baryon density dependent mean-field potential.
Since $v_3\{\Psi_1\}$ is only present for baryons in the data, it appears that the potential affects only baryons. Hence the presence of $v_3\{\Psi_1\}$ produced by a mean field potential acting only on baryons lends support to the conclusions of Ref. \cite{STAR:2021yiu} that the collisions studied here are in the hadronic phase where protons and neutrons are the dominant degrees of freedom and not the partonic phase.

\subsection{Comparison to HADES results at $\sqrt{s_{NN}} = 2.4$ GeV}
In order to make a comparison with HADES results at $\sqrt{s_{NN}} = 2.4$ GeV we employed the same selection of criteria for protons (20-30\% centrality and $1.0<p_{\mathrm{T}}<1.5$ GeV/$c$) \cite{HADES:2020lob} and extracted slope measurements from both energies. Using these selection criteria we replicated the plot from Fig. \ref{fig:prSymmetric} fitting it in the range $-0.5<y_{\mathrm{c.m.}}<0.5$ with the same equation as before --- $y=ax+bx^3$. Our resulting slope measurement was $dv_3/dy = -0.053 \pm 0.004$. We then applied the same fit to the data published by HADES for the proton $v_3\{\Psi_1\}$ which yielded a slope measurement of $dv_3/dy = -0.243 \pm 0.010$. This is about a factor of five larger than the value we obtained at $\sqrt{s_{NN}} = 3.0$ GeV. 

This may imply that the effect of the mean field is considerably stronger, even though the center-of-mass energy is lower by only 0.6 GeV. However, we must be cautious in that the center-of-mass energy may not be the relevant parameter. It is worth noting that, while the center-of-mass energy is only different by about 15\%, the available kinetic energy is about a factor of two more at $\sqrt{s_{NN}} = 3.0$ GeV. These observations prompt a need for further investigation into the physics that underlie these results to clarify and explain these disparities.

\section{\label{sec:level6}Summary }
$v_3\{\Psi_1\}$ correlated with the reaction plane has been measured in Au+Au collisions at 3 GeV in the Beam Energy Scan II at RHIC. The signal is seen predominantly for baryons (protons). $v_3\{\Psi_1\}$ values show a negative slope as a function of $y_{\mathrm{c.m.}}$, opposite to that of $v_1$ at this energy. Its magnitude is larger at higher rapidity, increases as one goes to higher $p_{\mathrm{T}}$ and towards more peripheral collisions. 

The development of $v_3\{\Psi_1\}$ is controlled by two key ingredients: the first is an appropriate geometry dictated by stopping, the passing time of the spectators, and the expansion of the fireball; the second is a potential in a responsive medium that drives the collective motion of particles. When compared with two theoretical models, JAM and SMASH, our data suggest that the required triangular geometry is a result of the dynamics of the collision, primarily from the nucleon-nucleon cross section, the energy, and the nuclear thickness which essentially bifurcates the nucleus when observed from one side in rapidity (stopping). 
In the models, JAM and SMASH, the force propelling the flow is provided by a potential which is a function of baryon density. Interestingly, it seems to affect only baryons and not mesons, at least within the statistical precision allowed by our present data set. In the models, any effects on mesons originate from the decays of excited baryons, or scattering, i.e. cascade type reactions. This, together with the fact that our data is consistent with these models, suggests that the medium is not in a partonic state and hence is below the QCD phase transition, consistent with the conclusions of previous STAR publications at $\sqrt{s_{NN}}=3$ GeV \cite{STAR:2021yiu, STAR:2021ozh}. 

A comparison with the HADES proton data indicates that the $v_3\{\Psi_1\}$ developed at $\sqrt{s_{NN}}=3$ GeV is much smaller than that at $\sqrt{s_{NN}}=2.4$ GeV. This does not necessarily mean that the mean field itself is stronger, in fact the parameters of the potential are probably similar as the energy increases, until a major change, such as a phase transition occurs. It is very likely that the three-dimensional geometry of the collision region is different at the two energies, for instance because stopping is different; hence this would change the effect of the potential on the motion of the medium.

Future data sets at higher energies should yield more information on the potential and its efficacy, as these will modify both the medium, particularly when transitioning through the phase change, and the geometry due to the decrease in stopping. The $v_3\{\Psi_1\}$ studied here can be used as a particularly sensitive probe to examine the validity of a model in which a mean field is used to describe the data, since $v_3\{\Psi_1\}$ does not seem to be developed from a cascade model. Detailed comparisons can begin to determine the form of the potential, although certain aspects such as the definition of centrality in the models should be revised to be more reflective of the definitions used in the data. This was not done here to keep the comparison simple. Before multiplicity is used in the models to define centrality, a careful comparison of the particle multiplicities, rapidity, and $p_{\mathrm{T}}$ distributions must be performed. 

An upcoming data sample with more than 5 times more statistics should yield more information, particularly for the $K^-$, and to verify whether the effect is exclusive to baryons, or whether there is a small effect for mesons. If the effect is only present for baryons, the disappearance of $v_3\{\Psi_1\}$ at higher energies may provide a signature for a situation in which baryons are no longer dominant and other degrees of freedom, e.g. constituent quarks, become important. In addition, the new data set will have increased PID capabilities as a new TOF detector (eTOF) which will extend time-of-flight coverage to $\eta=1.5$ (on the negative rapidity side using the coordinate system used in this analysis) was installed during data taking\cite{STAR:2016gpu}. Finally, we will be able to study this effect over the entire energy range covered by BES II. 

\begin{acknowledgments}
We thank the RHIC Operations Group and RCF at BNL, the NERSC Center at LBNL, and the Open Science Grid consortium for providing resources and support.  This work was supported in part by the Office of Nuclear Physics within the U.S. DOE Office of Science, the U.S. National Science Foundation, National Natural Science Foundation of China, Chinese Academy of Science, the Ministry of Science and Technology of China and the Chinese Ministry of Education, the Higher Education Sprout Project by Ministry of Education at NCKU, the National Research Foundation of Korea, Czech Science Foundation and Ministry of Education, Youth and Sports of the Czech Republic, Hungarian National Research, Development and Innovation Office, New National Excellency Programme of the Hungarian Ministry of Human Capacities, Department of Atomic Energy and Department of Science and Technology of the Government of India, the National Science Centre and WUT ID-UB of Poland, the Ministry of Science, Education and Sports of the Republic of Croatia, German Bundesministerium f\"ur Bildung, Wissenschaft, Forschung and Technologie (BMBF), Helmholtz Association, Ministry of Education, Culture, Sports, Science, and Technology (MEXT), Japan Society for the Promotion of Science (JSPS) and Agencia Nacional de Investigaci\'on y Desarrollo (ANID) of Chile.
\end{acknowledgments}

\bibliography{star_v3_3gev}
\end{document}


\preprint{APS/123-QED}

\title{Reaction plane correlated triangular flow in Au+Au collisions at $\mathbf{\sqrt{s_{\textrm{NN}}}=3}$ GeV}

\author{The STAR Collaboration}

\collaboration{STAR Collaboration}

\date{September 13, 2023}

\maketitle

\section{Event Plane Resolution}
\begin{figure}[!h]
    \centering
    \includegraphics[scale=0.32]{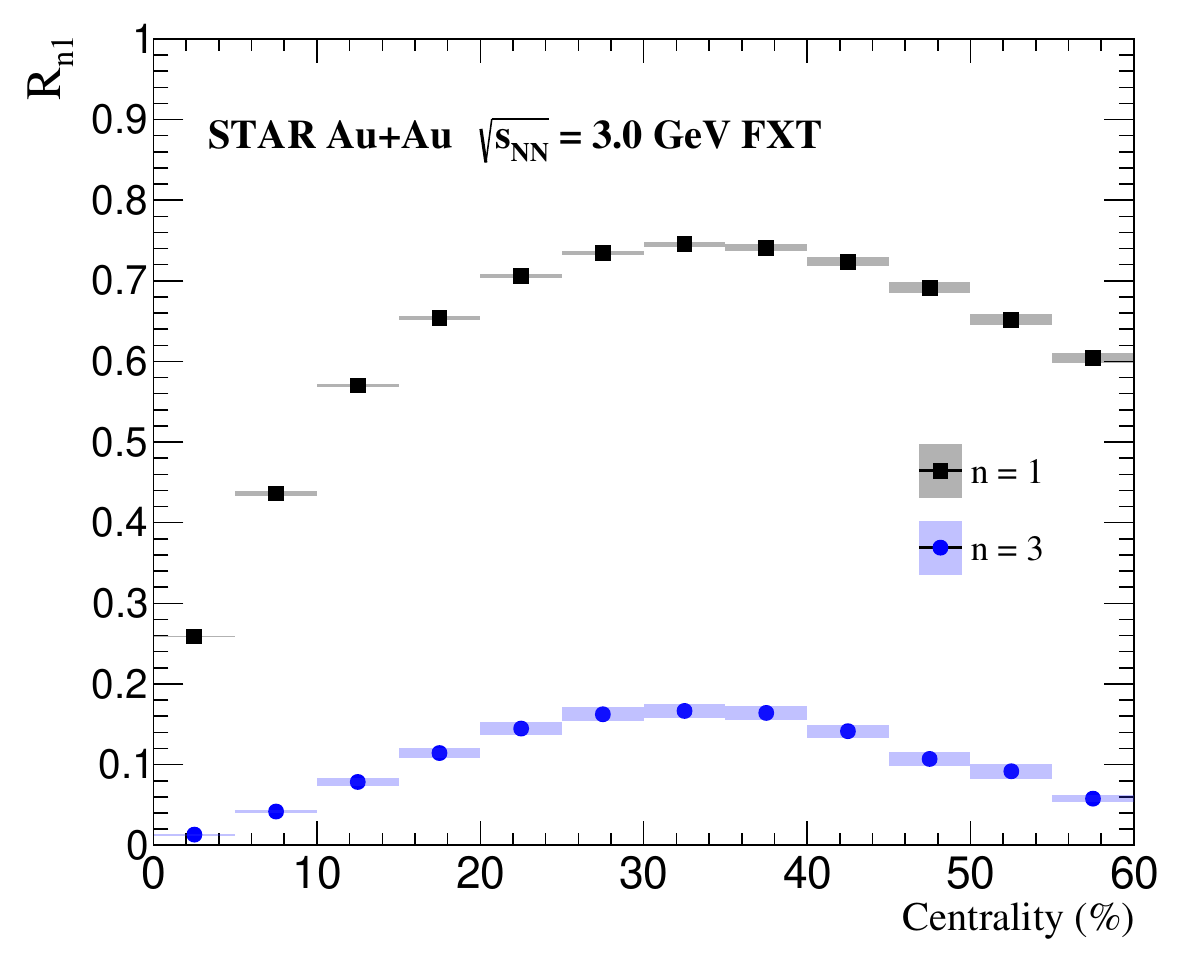}
    \caption{Event plane resolution for $v_1\{\Psi_1\}$ and $v_3\{\Psi_1\}$ as a function of centrality from $\sqrt{s_{\textrm{NN}}}=3$ GeV Au+Au collisions at STAR. The points are the average of the resolutions from the three configurations discussed in the main text, and the systematic uncertainties are the maximum difference between the configurations and the average (taken as a symmetric uncertainty in the opposite direction as well). Vertical lines are statistical uncertainties and shaded boxes are systematic uncertainties.}
    \label{fig:resolutions1and3}
\end{figure}

As a sanity check that our event plane resolutions for triangular flow using the first-order event plane ($v_3\{\Psi_1\}$) are reasonable, the event plane resolutions for directed flow ($v_1\{\Psi_1\}$) were also calculated. Figure \ref{fig:resolutions1and3} shows a comparison between the event plane resolutions for $v_1\{\Psi_1\}$ and $v_3\{\Psi_1\}$, where both sets were calculated with this analysis. The systematic uncertainties for both sets were calculated the same way as described in the main text. It was found that the first order resolutions are consistent with Ref. \cite{STAR:2021yiu}.

\bibliography{star_v3_3gev}